\newcount\style \style=1
\ifodd\style
  \documentstyle[12pt,aaspp4,psfig]{article}
\else
  \documentstyle[12pt,aasms4]{article}
\fi

\def\om{\Omega_{\rm p}}
\def\len{a_B}
\def\peak{R_{m=2,\rm max}}
\def\aqm{\cal A}
\def\lag{D_L}
\def\s{{\cal R}}
\def\n936{NGC 936}
\def\etal{{\it et al.}}
\def\eg{{\it e.g.}}
\def\etc{{\it etc.}}
\def\ie{{\it i.e.}}
\def\cf{{\it cf.}}
\def\spose#1{\hbox to 0pt{#1\hss}}
\def\gtsim{\mathrel{\spose{\lower.5ex \hbox{$\mathchar"218$}}
     \raise.4ex\hbox{$\mathchar"13E$}}}
\def\ltsim{\mathrel{\spose{\lower.5ex\hbox{$\mathchar"218$}}
     \raise.4ex\hbox{$\mathchar"13C$}}}
\def\nd{\nodata}
\def\bt#1{$#1(1-\epsilon)^2$}

\def\pmb#1{\setbox0=\hbox{$#1$}
  \kern-0.25em\copy0\kern-\wd0
  \kern.05em\copy0\kern-\wd0
  \kern-0.025em\raise.0433em\box0}
\def\bv{\;\pmb{\mit v}}
\def\bx{\;\pmb{\mit x}}

\slugcomment{\it Revised version to appear in the Astrophysical Journal}

\begin{document}

\title{Constraints from Dynamical Friction \\
on the Dark Matter Content of Barred Galaxies}
\author{Victor P. Debattista\altaffilmark1 and J. A. Sellwood}
\affil{Department of Physics and Astronomy, Rutgers, The State University of 
New Jersey, \\
136 Frelinghuysen Road, Piscataway, NJ 08854-8019}
\affil{debattis@astro.unibas.ch, sellwood@physics.rutgers.edu}
\altaffiltext{1}{Current address: Astronomisches Institut, Universit\"at 
Basel, Venusstrasse 7, CH-4102 Binningen, Switzerland}

\begin{abstract}
We show that bars in galaxy models having halos of moderate density and a variety of velocity distributions all experience a strong drag from dynamical friction unless the halo has large angular momentum in the same sense as the disk.  The frictional drag decreases the bar pattern speed, driving the corotation point out to distances well in excess of those estimated in barred galaxies.  The halo angular momentum required to avoid strong braking is unrealistically large, even when rotation is confined to the inner halo only.  We conclude, therefore, that bars are able to maintain their observed high pattern speeds only if the halo has a central density low enough for the disk to provide most of the central attraction in the inner galaxy.  We present evidence that this conclusion holds for all bright galaxies.

\end{abstract}

\section{Introduction}

The flatness of disk galaxy rotation curves, particularly outside the optical radius, is generally interpreted as evidence that they are embedded in massive dark matter (DM) halos.  Since the mass-to-light ratio of the visible material is uncertain, the central DM density is also uncertain; even the best determined 1-D rotation curve is consistent with almost any disk mass up to a maximum that does not over-fit the inner part (van Albada \etal\ 1985).  This degeneracy introduces a serious uncertainty into studies of galaxy formation and evolution.  Here we present an argument that DM halos have the lowest possible central density consistent with not being hollow.  Following van Albada \& Sancisi (1986), we refer to such minimum halo models as ``maximum disks.''

Strong bars are seen in optical images of roughly 30\% of all disk galaxies (Sellwood \& Wilkinson 1993) and this fraction rises to over 50\% in the near IR (Eskridge \etal\ 2000).  Their presence makes them ideal probes of the dynamics of the central regions.  The rate of rotation of a bar can be characterized by the ratio $\s = \lag/\len$, where $\lag$ is the corotation radius and $\len$ the bar semi-major axis.  More precisely, $\lag$ is the distance from the center to the Lagrange point on the bar major axis where the gravitational attraction balances the centripetal acceleration in the frame rotating with the bar.  Theoretical arguments (Contopoulos 1980) require $\s>1$, and there is a prejudice that $\s \gtsim 1$ (see Sellwood \& Wilkinson 1993 for a review).  We describe bars for which $1 \le \s \le 1.4$, \ie\ those for which corotation is not far beyond the bar's end, as fast.

The number of barred galaxies with measured $\s$ is still quite small (Elmegreen 1996), since it requires knowledge of the bar's pattern speed, $\om$, which is hard to determine.  Tremaine \& Weinberg (1984a) show that $\om$ can be measured directly from observations of a tracer component that satisfies the continuity equation.  To date, their method has been applied to just two galaxies: Merrifield \& Kuijken (1995, see also Kent 1987) find $\s = 1.4 \pm 0.3$ for NGC 936 while Gerssen \etal\ (1999) find $\s = 1.15^{+0.38}_{-0.23}$ for NGC 4596.  A third case should be completed soon (Debattista \& Williams 2000).  A less direct, but probably reliable, determination of $\s$ has been made for those few galaxies in which high spatial resolution, 2-D gas kinematics have been modeled.  Three cases are: $\s = 1.3$ in NGC 1365 (Lindblad \etal\ 1996), $\s = 1.3$ for NGC 1300 (Lindblad \& Kristen 1996), and $\s = 1.2$ for NGC~4123 (Weiner \etal\ 2000).  A more general argument can be made on the basis of the shapes of dust lanes that $\s = 1.2 \pm 0.2$ (van Albada \& Sanders 1982; Athanassoula 1992).  Other methods to determine $\s$ rely on identifying resonances, such as rings in the disk (\eg\ Buta 1986; Buta \& Combes 1996), but are less reliable.  Thus the meager reliable measurements are mostly for galaxies of earlier type, but all indicate bars are fast; this conclusion becomes much stronger, and can be extended to later Hubble types, if the dust lane argument holds.

Weinberg (1985) predicted that dynamical friction should brake the rotation rate of a bar on a time scale short compared with the ages of galaxies if a substantial density of dark matter is present in the region of the bar.  In a previous paper (Debattista \& Sellwood 1998, hereafter Paper I), we confirmed this prediction for non-rotating halos with isotropic velocity distributions and concluded that DM halos must have low central densities if bars are to remain as fast as those observed.  Here we extend this result to halos with anisotropic velocity dispersion tensors both with and without net rotation and show that it is not significantly altered for any reasonable velocity distribution of the dark matter (cf.\ Tremaine \& Ostriker 1999).

\section{Methods}

We present fully self-consistent, 3-D simulations of barred disks embedded in live halos.  Our simulations start from axisymmetric disk and halo models which are designed to be unstable to the formation of bars.

We adopt the Kuz'min-Toomre disk (Kuz'min 1956; Toomre 1963; Binney \& Tremaine 1987, \S2.2.1) because its density drops off less steeply than an exponential of the same scale length.  We give the disk a Gaussian density profile in the vertical direction with a scale height $z_{\rm d}$ and sharply truncate it at some radius $R_{\rm t}$.  Thus the disk density in cylindrical coordinates is
\begin{equation}
\rho(R,z) = \cases{ \displaystyle 
  \frac{ f_{\rm d} M_{\rm unit} } {(2 \pi)^{3/2} z_{\rm d} R_{\rm d}^2} \;
  \frac{\exp[{-\frac{1}{2} (z/z_{\rm d})^2}]} {[1 + (R/R_{\rm d})^2]^{3/2}} &
      $R \leq R_{\rm t}$ \cr
     \cr
  0 & $R > R_{\rm t}$ \cr},
\end{equation}
where $R_{\rm d}$ is the length scale of the disk and $f_{\rm d}$ is the fraction of mass in the disk.

We set up the initial halo from a distribution function (DF) that is computed to be in equilibrium in the presence of the adopted disk, as described in Appendix A.  We adopt a {\it lowered\/} polytrope form for the DF:
\begin{equation}
f(\bx,\bv) = C {\cal F}(E) = C \left\{ \left[-2 E(\bx,\bv) \right]^{n - \frac{3}{2}} - \left[-2E_{\rm max}\right]^{n - \frac{3}{2}} \right\}
\label{eqn:df}
\end{equation}
where $C$ is a normalization constant and $n$ and $E_{\rm max}$ are free parameters.  We emphasize that the halos generated this way are not polytropes; in particular, the density reaches zero at a finite radius for all $n > \frac{3}{2}$, whereas true polytropes of index $\geq 5$ have non-zero density everywhere.  By setting $E_{\rm max}$ to the potential energy in the plane of the disk at some radius within the grid, we guarantee that no particles are initially on orbits that would take them off the grid.  When $\cal F$ is a function if $E$ only, the DF is isotropic and the halo almost spherical, but in later sections we make $\cal F$ a function of a combination of $E$ and $J_z$ which leads to anisotropic DFs and spheroidal density distributions.  The procedure for selection of particles from a DF is described in Appendix B.  We set some of our halos into rotation by flipping the sign of $L_z$ for some fraction of the halo particles, as described in Appendix C.  In some cases, we flip particles in the inner part of the halo only, with a rule (equation \ref{eq:flipf}) that depends on energy in such a way that rotation declines continuously to zero near some (spherical) radius $r_{\rm rot}$.

We give the disk particles some random velocities with a radial dispersion, $\sigma_u$, set to yield a constant Toomre $Q$, neglecting any corrections arising from disk thickness and force softening.  We then use the epicyclic approximation to set the azimuthal velocity dispersion, $\sigma_v$ and the equation for asymmetric drift to set the mean orbital speed.  (This equation sometimes has no solution at small radii, particularly for large $Q$, in which case, we reduce $\sigma_u$.)  Finally, we set the vertical velocity dispersion, $\sigma_w$, from the 1-D vertical Jeans equation.  This disk set-up procedure is approximate, particularly for larger $Q$, but in practice the disk quickly adjusts to an equilibrium.

We have generally chosen $Q=0.05$ initially in order to hasten the formation of the bar, since we are here most interested in the evolution after this event.  Models run with higher initial $Q$ are not qualitatively different; even though the bars are initially weaker and buckle out of the plane more, they end up somewhat longer and are braked to an even greater extent.  The evolution is not significantly affected by changes to the truncation radius or by doubling or halving the initial thickness of the disk.

We imposed an initial three-fold symmetry on the models by replicating particles in sets of six in such a way as to ensure that the total momentum and components of the total angular momenta about the $x$- and $y$-axes (the $z$-axis being the symmetry axis) were all zero.  This prevented the model from rotating or drifting relative to our grid, which could lead to excessive and asymmetric loss of particles from the grid in our very long runs.

We adopt units in which $G = M_{\rm unit} = R_{\rm d} = 1$, where $G$ is Newton's constant.  The total of the disk ($M_{\rm d} = f_{\rm d}M_{\rm tot}$) and halo ($M_{\rm h}$) masses is $M_{\rm tot} = [ 1 - ( 1 + R_{\rm t}^2 / R_{\rm d}^2 )^{-1/2} ] \; M_{\rm unit}$.  Our unit of time is therefore $(R_{\rm d}^3/GM_{\rm unit})^{1/2}$, and our adopted time step is 0.05 in this unit.

We employ the 3-D Cartesian particle-mesh code described in Sellwood \& Merritt (1994), which uses an efficient FFT-based Poisson solver (James 1977).  This choice of code does limit us to computing the evolution within a fixed volume, and since we wish to retain reasonable spatial resolution within the disk, we have generally been forced to bound our halos at a small radius.  We find this a reasonable price to pay, since \eg\ treecode simulations of models with the same number of particles would run $\sim 100$ times more slowly (Sellwood 1997) which would preclude the extensive exploration of parameter space we report here.  Even the two simulations reported in \S\ref{maxdisk}, which use larger grids to permit more extensive halos, run only $\sim12$ times more slowly than our standard grid, and are therefore still decisively less expensive than a treecode.  These performance comparisons are all based on a single processor general purpose workstation; the advantage of a grid code would be even greater on parallel computers where the field evaluation is more easily optimized.  Athanassoula \etal\ (1998) give performance comparisons between the grid code and a machine with special-purpose hardware (GRAPE3), which clearly must depend upon the workstation adopted for comparison, but from their Figure 3, it can be seen that the grid code is competitive with their GRAPE3 machine for the grid size and particle number we employ.

\begin{table}[t]
\vbox{\hfil
\begin{tabular}{ r||l }
 Halo type & Lowered polytrope, $n=3$ \\ 
 Disk type & Kuz'min-Toomre \\ 
 Disk scale length, $R_{\rm d}$ & 5 mesh spaces \\ 
 Disk scale height, $z_{\rm d}$ & $0.1R_{\rm d}$ \\ 
 Disk truncation radius, $R_{t}$ & $4 R_{\rm d}$ \\ 
 Disk mass fraction, $f_{\rm d}$ & $0.3$ \\ 
 Disk Toomre $Q$ & 0.05 \\ 
 Halo truncation radius, $r_{\rm h}$ & $12.6 R_{\rm d}$ \\ 
 Grid size and type & $129 \times 129 \times 129$ Cartesian\\ 
 Number of particles, $N$ & 300,000 (equal mass) \\ 
 Time step & $0.05 \ (R_{\rm d}^3/GM_{\rm unit})^{1/2}$ \\
\end{tabular}
\hfil}
\caption{Properties and parameters in the canonical simulation.  These are default values for all simulations reported in this paper, except where noted otherwise.}
\label{tab:numpars}
\end{table}

The numerical parameters in the simulations we report here are summarized in Table \ref{tab:numpars}.  We have checked that our main conclusions are insensitive to changes in particle number, to an increase in spatial resolution or the method to determine the gravitational field, \etc\ ~In particular, the evolution of the pattern speed and bar length on finer grids, or with different geometry tracked that on our standard grid pretty well, see Debattista (1998) for details.  We repeated three simulations with identical physical properties but different random seeds in the generation of particles and report the results in Figure~\ref{svslhalo}.  Some particles were lost from the grid (typically no more than 3\% in the longest runs) almost all of which were halo particles.  Naturally, these weakly bound particles carried away more than their fair share of angular momentum, which decreased by as much as 5\% in the worst case.  A test on a larger grid showed that the evolution is imperceptibly affected by this loss.

\subsection{Pattern speed and Lagrange point}
\label{lagpt}

In order to determine the bar pattern speed, $\om$, and other properties of our models, we need a well-defined center about which to perform an expansion.  Despite our careful set up, the center of our $N$-body system wanders from the center of our grid.  Following McGlynn (1984), we define the function:
\begin{equation}
\omega_k(x_0,y_0,z_0) \ = \ \sum_{j=1}^N \ \left[ (x_j - x_0)^2 + (y_j - y_0)^2 +  (z_j - z_0)^2 \right]^k
\end{equation}
where $(x_j,y_j,z_j)$ are the coordinates of the $j$th particle and $(x_0,y_0,z_0)$ those of an expansion center.  Minimizing $\omega_k$ with respect to $x_0$, $y_0$ and $z_0$ gives a centroid for the system.  Setting $k = 1$ yields the center of mass, distant particles are weighted more when $k > 1$ while $k < 1$ gives a centroid more sensitive to small scales.  We adopt $k = \frac{1}{2}$, which removes the dipole moment.  We determine the centroid for the total system of particles, disk and halo; the disk and halo centroids typically differed significantly only during the brief interval when the bar buckled, when the position angle of the bar was anyway not well defined.  We obtain an improved estimate of the centroid position from a single Newton iteration of its old value every 20 steps, immediately before each analysis step.  We have verified, at a few selected times, that further refinement results in a change in the position of the centroid by $\ltsim 0.02 R_{\rm d}$.

We expand the instantaneous distribution of disk particles in logarithmic spirals as
\begin{equation}
\label{logspi}
A(m,\gamma,t) = \frac{1}{N_{\rm d}} \sum_{j=1}^{N_{\rm d}} \exp im\left( \varphi_j + \tan\gamma \ln R_j\right),
\end{equation}
where $\gamma$ is the angle between the radius vector and the ridge line of the spiral.  Here, $(R_j,\varphi_j)$ are cylindrical polar coordinates (with respect to the centroid) of the $j$-th particle at time $t$.  The $m = 2$, $\gamma = 0$ term of this expansion gives the phase and amplitude of the bar
\begin{equation}
A(2,0,t) = \aqm {\rm e^{2i\varphi}},
\end{equation}
where $\aqm$ is the bar amplitude.  We calculate the {\it monotonic\/} angular displacement of the bar, $\phi(t)$, from $\varphi$.  We estimate its derivative, $\om(t) \equiv \dot\phi(t)$, by fitting a straight line to 25 consecutive data points centered at $t$, which smoothes out rapid fluctuations in $\om$ and yields an error estimate from the standard error in this linear fit.

Having determined $\om$, we are in a position to calculate $\lag$.  We determine the effective force along the bar major axis in the disk plane (we average the gravitational force from both sides of the center, and use the {\it instantaneous\/} value of $\om$); $\lag$ is the distance from the center at which the net force passes through zero.  We use the uncertainty in $\om$ to determine that in $\lag$ directly.  This procedure is superior to determining the radius at which $R \om$ intersects an axisymmetric rotation curve; we have found that $\lag$ is larger by more than $\sim5$\% for strong bars when $\s \simeq 1$.

It should be noted that the value of $\lag$ is affected by the rotation curve shape.  As the bar slows, the distance from the center to the Lagrange point increases more slowly when the rotation curve declines than it would do if the rotation curve were flat.  Since, in the large majority of our simulations the rotation curve does in fact drop continuously from the maximum in the disk, our reported values of both $\lag$ and $\s$ are underestimates of the values they would have in more realistic models with flat rotation curves (see \S5).

\subsection{Bar semi-major axis}

A bar is a straight bisymmetric distortion in the density of a disk; even Fourier components of the surface density are therefore ideal for tracing the extent of the bar.  Thus, for example, Lindblad \etal\ (1996) and Lindblad \& Kristen (1996) used the phase variations of the even Fourier components to determine the lengths of the bars in NGC 1365 and NGC 1300 respectively.  Here we adopt a similar approach for determining the semi-major axis, $\len$, of the bars in our simulations, using only the $m=2$ Fourier component.  This is not always an easy measurement, particularly when the disk possesses $m=2$ disturbances other than the bar, such as spirals, rings surrounding the bar, and outer oval distortions.

We divide the disk into radial bins of fixed width $0.16\,R_{\rm d}$ and determine the amplitude and phase of the second sectoral ($m=2$) harmonic from the particle positions within each annular bin.  We estimate the errors $\sigma_{\rm amp}$ and $\sigma_{\rm phs}$ using Monte-Carlo measurements of the phase and amplitude from synthetic samples of particles drawn from a distribution with a given $m=2$ amplitude.  Fitting these data with a 2-D spline then yielded interpolation formulae, giving the uncertainties for the number of particles in each annulus and the measured $m=2$ amplitude.

Spirals are the easiest to distinguish, since they generally have a different pattern speed from the bar.  Thus at a fixed radius, the peak of a spiral's azimuthal position is usually different from that of the bar.  When the inner part of the spiral lines up with the bar, however, there is no way to distinguish between it and the bar; measurements of $\len$ therefore fluctuate at the beat frequency of the bar and spiral pattern speeds (this same beat frequency can be seen in measurements of $\om$).  This problem becomes less severe as the evolution proceeds, because a rising $Q$ causes the spirals to weaken.

Oval outer disks are often perpendicular to the bar (in this sense, this outer region can be considered as an outer ring of the type R$_1$ in Buta's [1986] classification).  The number of particles in these ovals is often low, and the error bars associated with their position angle correspondingly large enough to confuse the measurement of $\len$, especially late in the runs when the surface density just beyond the bar's end has been depleted.  To counteract this tendency, we did not include radial bins in which the error in position angle is greater than $\sim 80^\circ$.  An example of a mildly oval outer disk can be seen in Figure \ref{pntplt}.

At later stages in the disk's evolution, the spirals often settle to form a ring around the bar, as can be seen in Figure \ref{pntplt}.  Their position (just outside the bar) and orientation (usually aligned with the bar) suggest they are inner rings (Athanassoula \& Bosma 1985; Buta 1986; Buta \& Combes 1996).\footnote{The presence of these rings in our simulations has important ramifications for the theory of ring formation and interpretation, since rings are often considered to be gas phenomena, but our collisionless simulations have no gas.}  Since these elliptical rings mostly line up with the bar, they constitute an additional $m=2$ component locked at the bar's pattern speed which further complicates identification of the bar end.  When the disk is sub-divided into annuli and the $m=2$ amplitude plotted as a function of radius, the ring appears as an upwards bump in the smooth decrease of amplitude from the bar.  We adopt, as one estimate denoted ${\len}_1$, the last radial point at which this radially binned $m=2$ amplitude did not deviate from a linear decrease by more than its standard error, $\sigma_{\rm amp}$.

We obtain a second estimate, ${\len}_2$, from the phase of the $m=2$ moment of the disk binned as for the ${\len}_1$ measurement.  We estimate ${\len}_2$ as the radius of the outermost bin for which the phase is constant to within the standard error in that radial bin, $\sigma_{\rm phs}$.

We define $\len$ to be the simple average of ${\len}_1$ and ${\len}_2$.  In practice, ${\len}_1$ tended to underestimate our subjective visual impression of the bar semi-major axis (particularly at early times), while ${\len}_2$ tended to overestimate it.  We found that the average of these two quantities did a very good job of estimating $\len$.  We generously define the uncertainty in $\len$, which is not a formal error, as half the difference between ${\len}_1$ and ${\len}_2$.  Because of the formation of rings and ovals, this uncertainty tends to be largest at late times, and can be as large as several disk scale-lengths.

The uncertainty in $\lag$ is always much less than our generous estimates of the uncertainty in $\len$, which therefore dominates our quoted uncertainty in $\s$.

\subsection{Parameterizing rotation curve decompositions}

\label{sec:params}

One estimator of the relative contributions of the disk and halo to the central attraction is the parameter 
\begin{equation}
\eta \equiv \left. \left(\frac{V_{\rm c,disk}} {V_{\rm c,halo}} 
\right)^2\right|_{R_{\rm m}}
\end{equation}
where $V_{\rm c,disk}$ and $V_{\rm c,halo}$ are the circular velocities due to the disk and halo respectively.  In Paper I, we defined $\eta_{\rm exp}$ at $R_{\rm m,exp}$, the disk-plane radius at which $V_{\rm c,disk}$ would peak for an infinite, razor thin exponential disk, as appropriate for the model with the extensive halo; we then adopted $R_{\rm m} = 1.75R_{\rm d}$ as the nearest equivalent for the Kuz'min-Toomre disks.  Here, however, we define $\eta$ at the true disk maximum for the Kuz'min-Toomre disk, $R_{\rm m} = 1.41R_{\rm d}$, since we employ that disk in all the simulations reported here.  It should be noted that even though the rotation curve evolves as our simulations proceed, our values for $\eta$ are those of the initial model only.

Here we explore a wider range of models than in Paper I, with a greater variety of rotation curve shapes and find, not surprisingly, that a parameter which depends on the ratio of rotation velocities from disk and halo at a single radius does not correlate well with the degree of braking we observe in our simulations.  Furthermore, halo angular momentum changes the evolution of $\s$, so that $\eta$ is clearly an inadequate predictor of $\s$ even for fixed rotation curves.

A parameter which takes into account both the angular momentum in the halo and the full rotation curve might do a better job.  We define
\begin{equation}
\Gamma(r_0) \equiv \frac{\sum_{i, r < r_0} |J_{z,{\rm h}, i}| - J_{z,{\rm h}, i} } {\sum_{i, R<4R_{\rm d}} J_{z,{\rm d}, i} }
\end{equation}
where $J_{z,{\rm h}, i}$ is the angular momentum of the $i$-th halo particle about the symmetry axis, $r_0$ is some arbitrary (spherical) cutoff radius for the summation and $J_{z,{\rm d}, i}$ is the angular momentum of the $i$-th disk particle.  The quantity $\Gamma(r_0)$ is the difference between the maximum possible and the actual angular momentum content of the halo, expressed as a fraction of disk's angular momentum, and is zero for a maximally rotating halo.  It can be thought of as a measure of the capacity of the inner halo to accept angular momentum; it depends both on the halo mass distribution as well as its angular momentum content.

\section{Massive Halo Models}

\begin{deluxetable}{ccccccccccccc}
\tablewidth{0pt}
\tablecaption{All runs in this paper}
\scriptsize
\tablehead{
\colhead{Run\tablenotemark{a}}  & 
\colhead{$f_{\rm d}$}  & 
\colhead{$Q$}  & 
\colhead{$n$}  & 
\colhead{$\cal{H}$}  & 
%\colhead{$\frac{L_{z,{\rm h}}}{L_{z,\rm max}}$}  & 
\colhead{$\beta$}  & 
\colhead{$L_{z,{\rm h}}$ pars\tablenotemark{b}} &
%\alpha$ (s/l)}  \colhead{$d$}  & 
\colhead{$\eta$}  & 
\colhead{$\Gamma(3)$}  & 
\colhead{$\lag/\len(1000)$\tablenotemark{c}}  & 
\colhead{$\s_{\rm lmp}$\tablenotemark{d}}  & 
\colhead{$t({\rm lmp})$\tablenotemark{e}}  & 
\colhead{Comments}  }
\startdata

\nl
\cutinhead{{\bf Halo type: ${\cal F}(E)$}}

  *4 & 0.3 & 0.05 & 3 &   0.0 &  0.0 &      \nd & 2.0 & 0.63 & $6.6/(2.8\pm0.2)$ & $1.9\pm 0.5$ &2000& Canonical run   \nl  					                        
 20 & 0.3 & 0.05 & 3 & -0.98 &  0.0 & 34.97 (l) & 2.0 & 1.23 & $7.1/(2.0\pm0.2)$ & $3.5\pm 0.3$ &1000& \nd             \nl  					                        
 21 & 0.3 & 0.05 & 3 &  0.98 &  0.0 & 34.97 (l) & 2.0 & 0.02 & $4.2/(2.5\pm0.3)$ & $1.7\pm 0.2$ &1000& \nd             \nl  					                        
 22 & 0.3 & 0.05 & 3 &  0.33 &  0.0 &  0.74 (s) & 2.0 & 0.47 & $5.6/(3.0\pm0.5)$ & $1.9\pm 0.3$ &1000& \nd             \nl  					                        
 23 & 0.3 & 0.05 & 3 &  0.66 &  0.0 &  5.39 (l) & 2.0 & 0.27 & $4.9/(3.4\pm0.6)$ & $1.4\pm 0.3$ &1000& \nd             \nl  					                        
 24 & 0.3 & 0.05 & 3 & -0.33 &  0.0 &  0.74 (s) & 2.0 & 0.79 & $6.9/(2.6\pm0.2)$ & $2.6\pm 0.2$ &1000& \nd             \nl  					                        
 25 & 0.3 & 0.05 & 3 & -0.66 &  0.0 &  5.39 (l) & 2.0 & 0.98 & $5.6/(2.2\pm0.2)$ & $2.6\pm 0.2$ &1000& \nd             \nl  					                        
  7 & 0.3 & 1.0  & 3 &   0.0 &  0.0 &      \nd & 2.0 & 0.67 & $5.6/(3.0\pm 0.4)$ & $3.3\pm0.3$ &2500& \nd              \nl  
  8 & 0.3 & 1.5  & 3 &   0.0 &  0.0 &      \nd & 2.0 & 0.70 & $4.5/(3.0\pm 0.2)$ & $3.2\pm0.2$ &3250& \nd              \nl  
 28 & 0.3 & 0.05 & 3 &   0.0 &  0.0 &      \nd & 2.0 & 0.63 & $5.7/(2.3\pm0.2)$  & $1.8\pm 0.6$ &2000& Twin of Run 4   \nl  					                        
 33 & 0.3 & 0.05 & 3 &  0.98 &  0.0 & 34.97 (l)& 2.0 & 0.02 & $3.5/(2.5\pm0.6)$ & $1.7\pm 0.4$ &2000& Twin of Run 21  \nl  					                        
 34 & 0.3 & 0.05 & 3 & -0.98 &  0.0 & 34.97 (l)& 2.0 & 1.23 & $7.1/(2.4\pm0.4)$ & $3.3\pm 0.7$ &2000& Twin of Run 20  \nl  
 44 & 0.3 & 0.05 & 3 &   0.0 &  0.0 &      \nd & 2.0 & 0.63 & $6.5/(3.9\pm 1.0)$ & $2.5\pm0.4$ &1500& $z_{\rm d} = 0.2R_{\rm d}$ \nl  
 47 & 0.3 & 0.05 & 3 &   0.0 &  0.0 &      \nd & 2.0 & 0.63 & $6.3/(2.6\pm 0.2)$ & $2.4\pm0.1$ &1000& $z_{\rm d} = 0.05R_{\rm d}$ \nl

\cutinhead{{\bf Halo type: ${\cal F}(E + \beta J_z^2)$}}
 50 & 0.3 &  1.0 & 3 &   0.0 & -0.1 &       \nd & 2.4 & 0.89 & $6.9/(3.0\pm0.2)$ & $2.3\pm 0.1$ &1000&$257^2\times 129;r_{\rm h}=8R_{\rm d}$\nl  		                        
 54 & 0.3 &  1.0 & 3 &   1.0 & -0.1 &      full & 2.4 & 0.00 & $4.5/(3.4\pm0.6)$ & $1.4\pm 0.3$ &1250&$r_{\rm h}=8R_{\rm d}$\nl  				                        
 55 & 0.4 &  1.0 & 3 &   0.0 &  0.1 &       \nd & 1.4 & 0.50 & $5.6/(3.7\pm0.7)$ & $2.0\pm 0.2$ &1750&$R_{\rm t}=8R_{\rm d}$ \nl  					                        
 56 & 0.4 &  1.0 & 3 &   1.0 &  0.1 &      full & 1.4 & 0.00 & $4.4/(3.6\pm1.0)$ & $1.2\pm 0.4$ &1500&$R_{\rm t}=8R_{\rm d}$ \nl  
                                                 
\cutinhead{{\bf Halo type: ${\cal F}[E + \beta(E) J_z^2]$}}
123 & 0.3 &  1.0 & 3 &   0.0 & \bt{-0.2} &       \nd & 2.9 & 0.67 & $7.1/(3.0\pm0.2)$ & $2.4\pm 0.4$ &1250& \nd             \nl  					                        
124 & 0.3 &  1.0 & 3 &   1.0 & \bt{-0.2} &      full & 2.9 & 0.00 & $3.5/(2.6\pm0.4)$ & $1.3\pm 0.1$ &1250& \nd             \nl  					                        
137 & 0.3 &  1.0 & 3 &   0.04& \bt{-0.2} &       2.0 & 2.9 & 0.55 & $6.7/(3.3\pm0.2)$ & $2.1\pm 0.2$ &1250&$r_{\rm rot}=2R_{\rm d}$\nl  				                        
138 & 0.3 &  1.0 & 3 &   0.15& \bt{-0.2} &       1.5 & 2.9 & 0.28 & $4.9/(2.8\pm0.5)$ & $1.7\pm 0.3$ &1000&$r_{\rm rot}=4R_{\rm d}$\nl  				                        
139 & 0.3 &  1.0 & 3 &   0.50& \bt{-0.2} &       1.2 & 2.9 & 0.06 & $4.2/(2.4\pm0.2)$ & $1.6\pm 0.3$ &1250&$r_{\rm rot}=6R_{\rm d}$\nl  				                        
142 & 0.3 &  1.0 & 3 &   0.0 & \bt{-0.5} &       \nd & 4.8 & 0.67 & $5.8/(3.0\pm0.2)$ & $2.1\pm 0.1$ &2000& \nd             \nl  					                        
143 & 0.3 &  1.0 & 3 &   0.11& \bt{-0.5} &       1.5 & 4.8 & 0.33 & $3.3/(3.0\pm0.8)$ & $1.5\pm 0.3$ &2000&$r_{\rm rot}=4R_{\rm d}$\nl  				                        
144 & 0.3 &  1.0 & 3 &   1.0 & \bt{-0.5} &      full & 4.8 & 0.00 & $3.6/(2.8\pm0.3)$ & $1.3\pm 0.2$ &2000& \nd             \nl  					                        
145 & 0.3 &  1.0 & 3 &   0.03& \bt{-0.5} &       2.0 & 4.8 & 0.58 & $3.8/(2.2\pm0.2)$ & $1.6\pm 0.2$ &2000&$r_{\rm rot}=3R_{\rm d}$\nl  				                        
146 & 0.3 &  1.0 & 3 &   0.47& \bt{-0.5} &       1.2 & 4.8 & 0.07 & $3.5/(2.9\pm0.4)$ & $1.3\pm 0.1$ &2000&$r_{\rm rot}=7R_{\rm d}$\nl  				                        
147 & 0.3 &  1.0 & 3 &   0.04& \bt{-0.5} &       1.8 & 4.8 & 0.51 & $3.6/(2.6\pm0.2)$ & $2.3\pm 0.2$ &3500&$r_{\rm rot}=3R_{\rm d}$\nl  
125 & 0.3 &  1.0 & 3 &   1.0 & \bt{-0.1} &  full     & 2.4 & 0.00 & $3.7/(2.3\pm0.3)$ & $1.3\pm0.1$  &1250& \nd                \nl  
135 & 0.3 &  1.0 & 3 &   0.0 & \bt{-1.0} & \nd      & 10.0 & 0.66 & $3.1/(2.6\pm0.6)$ & $1.2\pm0.3$ &1000& \nd                 \nl  

\tablebreak

\cutinhead{{\bf Halo type: ${\cal F}(E)$ with other mass ratios and polytrope indices}}
 *59 & 0.4 & 0.05 & 3 &   0.0 &  0.0 &      \nd & 2.6 & 0.41 & $5.1/(3.2\pm0.4)$ & $1.6\pm 0.4$ &2000& \nd             \nl  					                        
 *60 & 0.5 & 0.05 & 3 &   0.0 &  0.0 &      \nd & 3.4 & 0.28 & $4.7/(2.9\pm0.9)$ & $1.3\pm 0.1$ &2000& \nd             \nl  					                        
 *61 & 0.6 & 0.05 & 3 &   0.0 &  0.0 &      \nd & 4.6 & 0.19 & $4.0/(2.9\pm0.9)$ & $1.2\pm 0.1$ &1500& \nd             \nl  					                        
 72 & 0.3 & 0.05 &1.6&   0.0 &  0.0 &       \nd & 3.4 & 0.54 & $3.7/(1.8\pm0.2)$ & $2.0\pm 0.5$ &2500&$R_{\rm t}=6R_{\rm d}$ \nl  					                        
 76 & 0.4 & 0.05 &1.6&   0.0 &  0.0 &       \nd & 4.8 & 0.36 & $5.5/(3.4\pm0.9)$ & $2.1\pm 0.3$ &2000&$N = 240K;R_{\rm t}=6R_{\rm d}$\nl  				                        
 80 & 0.5 & 0.05 &1.6&   0.0 &  0.0 &       \nd & 6.3 & 0.24 & $4.5/(3.4\pm0.6)$ & $1.8\pm 0.5$ &1750& \nd             \nl  					                        
 81 & 0.3 & 0.05 & 4 &   0.0 &  0.0 &       \nd & 1.0 & 0.71 & $5.1/(3.0\pm0.7)$ & $2.4\pm 0.1$ &2500& \nd             \nl  					                        
 83 & 0.4 & 0.05 & 4 &   0.0 &  0.0 &       \nd & 1.4 & 0.46 & $4.8/(2.9\pm0.2)$ & $1.4\pm 0.1$ &2250& \nd             \nl  					                        
140 & 0.4 & 0.05 & 4 &   0.0 &  0.0 &       \nd & 1.4 & 0.46 & $5.0/(3.2\pm0.7)$ & $1.4\pm 0.2$ &2000& Twin of run 83      \nl  
 87 & 0.5 & 0.05 & 4 &   0.0 &  0.0 &       \nd & 1.9 & 0.31 & $4.6/(3.3\pm0.8)$ & $1.3\pm 0.2$ &1750& \nd     \nl
 
\cutinhead{{\bf Large volume ($257^3$) runs; $N = 600K$; $r_{\rm h} = 25.2 R_{\rm d}$; $R_{\rm t} = 8 R_{\rm d}$. Halo type: ${\cal F}(E)$}}
 68 & 0.17& 0.05 & 2 &   0.0 &  0.0 &       \nd & 7.0 & 0.39 & $3.0/(2.6\pm0.2)$ & $1.6\pm 0.3$ &2000& Maximum disk run\nl  					                        
141 & 0.17& 0.05 & 3 &   0.0 &  0.0 &       \nd & 3.8 & 0.58 & $3.8/(2.4\pm0.2)$ & $2.0\pm 0.4$ &2000& Control run     \nl  

\enddata
\tablenotetext{a}{Runs also presented in Paper I are marked with an asterisk.}
\tablenotetext{b}{Parameter settings in generating halo angular momentum.  For isotropic halos, we report the value of $\alpha$ followed by (l) or (s) depending upon whether the angular momentum needed is large or small in equation \ref{eq:flipe}.  For varying anisotropy halos, we report the value of $d$.  An entry of ``full'' in the column means that all retrograde particles have been flipped to give the initial setup.}
\tablenotetext{c}{The value of $\lag/\len$ by $t=1000$, the minimum duration of all simulations.}
\tablenotetext{d}{The value of $\s$ at the last measured point, which is not always when the bar has finished slowing down.  All bars in our simulations form with $\s \simeq 1$.}
\tablenotetext{e}{The time at which $\s_{\rm lmp}$ is reported.  Note that one rotation at $R_{\rm d}$ takes 19 time units in the canonical run.}

\label{tab:allruns}
\end{deluxetable}

\ifodd\style
\begin{figure}[t]
\centerline{\psfig{figure=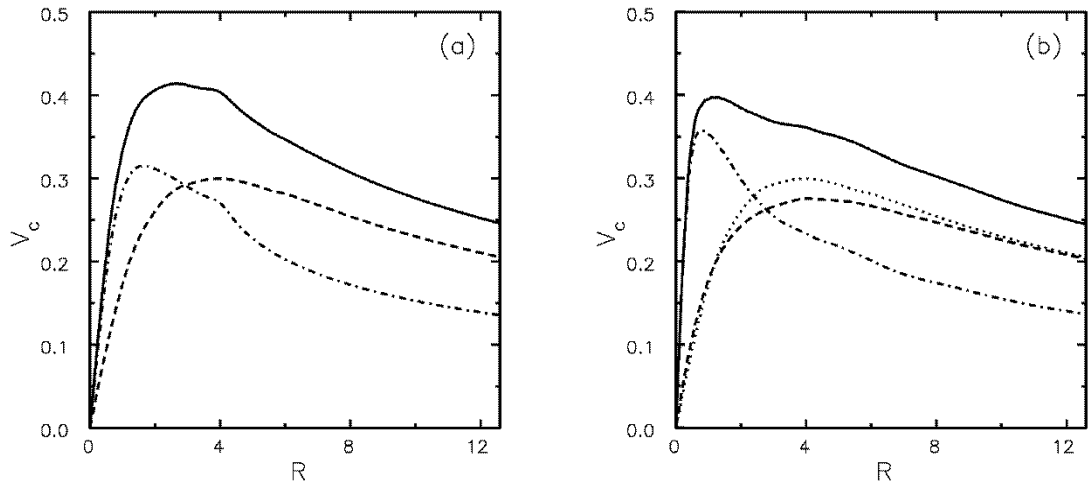,width=0.9\hsize,clip=}}
\caption{(a) The initial rotation curve of the canonical run.  The unbroken line is the total rotation curve, the dot-dashed line is the disk contribution and the dashed line is the halo contribution.  (b) The rotation curve of the axisymmetric (azimuthally shuffled) particle distribution towards the end of the simulation.  The extra (dotted) line shows the halo contribution at $t=0$ for comparison.  Note that angular momentum redistribution resulted in a more concentrated disk, and a slightly lower density halo.}
\label{4rotcurs}
\end{figure}
\fi

We begin by describing a set of experiments with disks embedded in moderately dense halos.  Halos of significantly greater density than we use here would inhibit the formation of the bar (Ostriker \& Peebles 1973; Toomre 1981).  We already demonstrated in Paper I that such a halo having an isotropic velocity distribution would brake the bar to an unacceptable extent.  Here we determine the extent to which braking is affected by giving the halo net angular momentum, or an anisotropic distribution of velocities, or both.

\ifodd\style
\begin{figure}[t]
\centerline{\psfig{figure=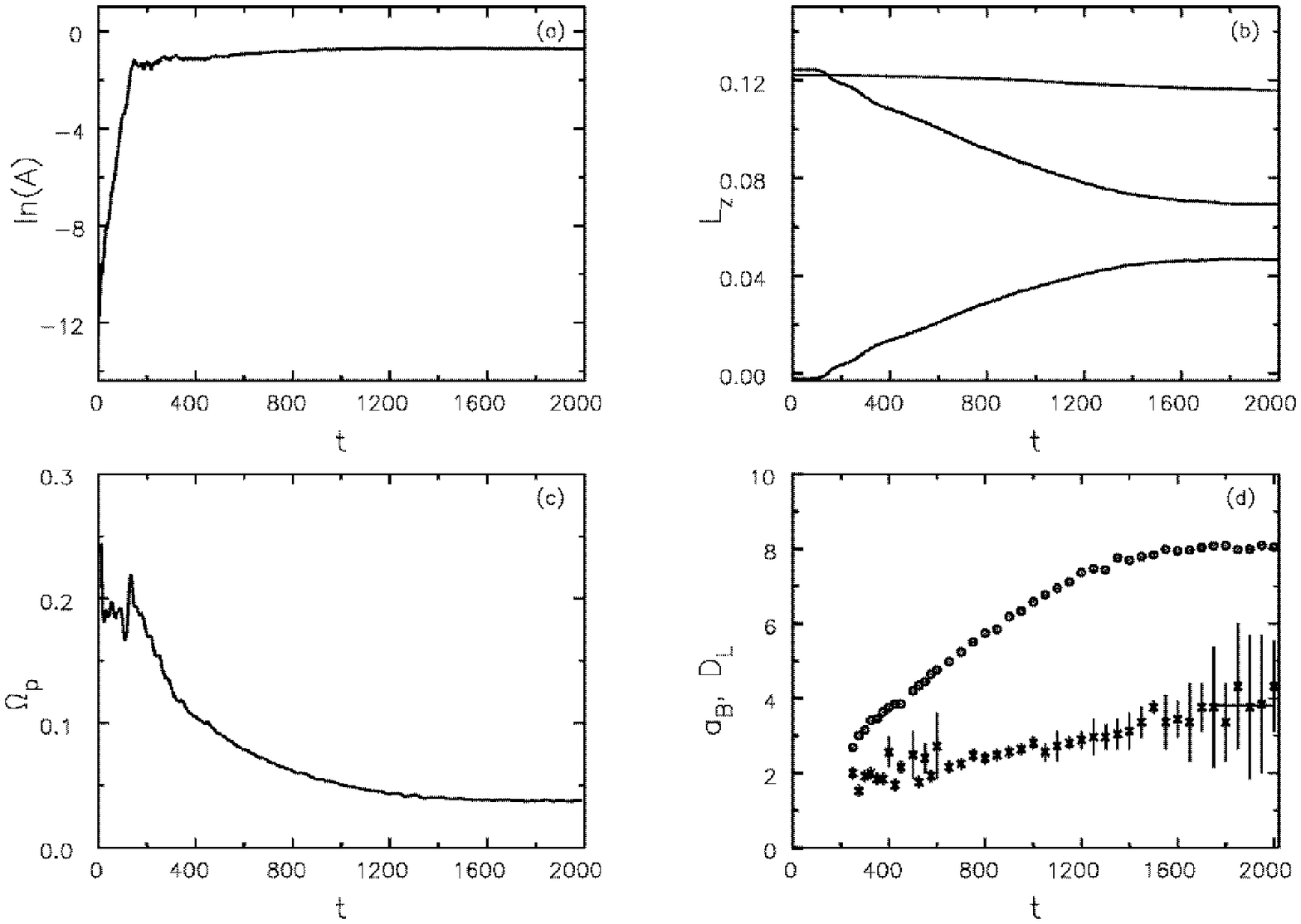,width=0.8\hsize,clip=}}
\caption{The evolution of the canonical run.  (a) The amplitude of the bar grows very rapidly, experiences a weak buckling around $t=200$ and then continues to grow slowly.  (b) The total (topmost), disk (middle) and halo (bottom) angular momentum.  Angular momentum is lost from the disk and gained by the halo, with a little carried off the grid by escaping particles.  (c) The pattern speed drops rapidly soon after the bar forms, but it reaches a constant value by $t \sim 1600$.  (d) $\len$ (crosses) and $\lag$ (circles).  The horizontal line joining the last 6 values of $\len$ shows the weighted average final value of $\len$ , which gives $\s = 2.1 \pm 0.2$.}
\label{4abcd}
\end{figure}
\fi

\ifodd\style
\begin{figure}[p]
\centerline{\psfig{figure=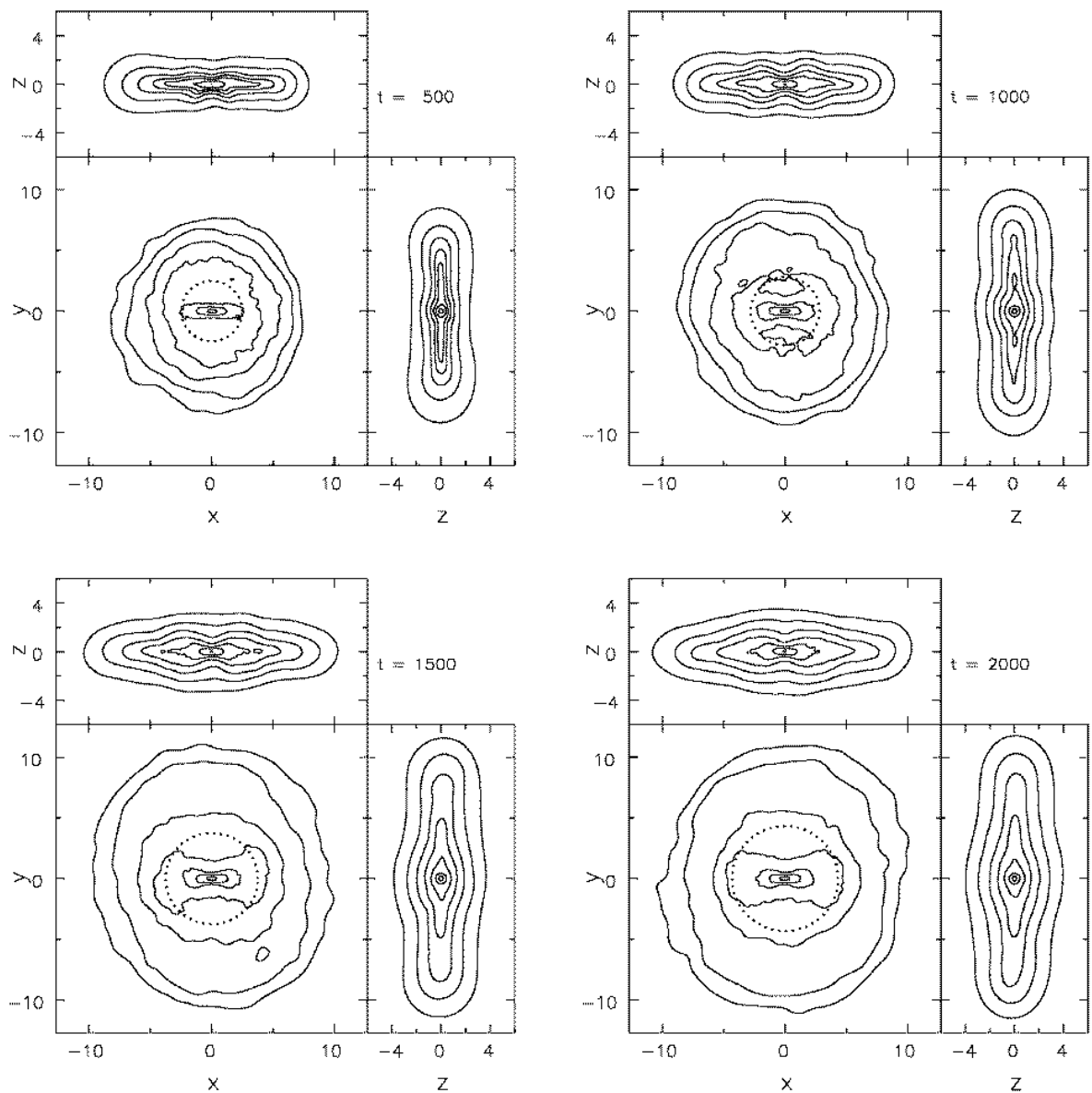,width=1.0\hsize,clip=}}
\caption{Contours of projected disk density in the canonical simulation at 
four instants.  The bar has been rotated into the $x$-axis.  Contours are 
logarithmically spaced.  The circle in the $(x,y)$ projection shows $\len = 
2.48$, 2.80, 3.76 \& 4.32 at $t=500$, 1000, 1500 \& 2000 respectively.  Note that the outer disk is distinctly oval, and the peanut-shaped isophotes in the edge-on view.}
\label{cont10000}
\end{figure}
\fi

\ifodd\style
\begin{figure}[t]
\centerline{\psfig{figure=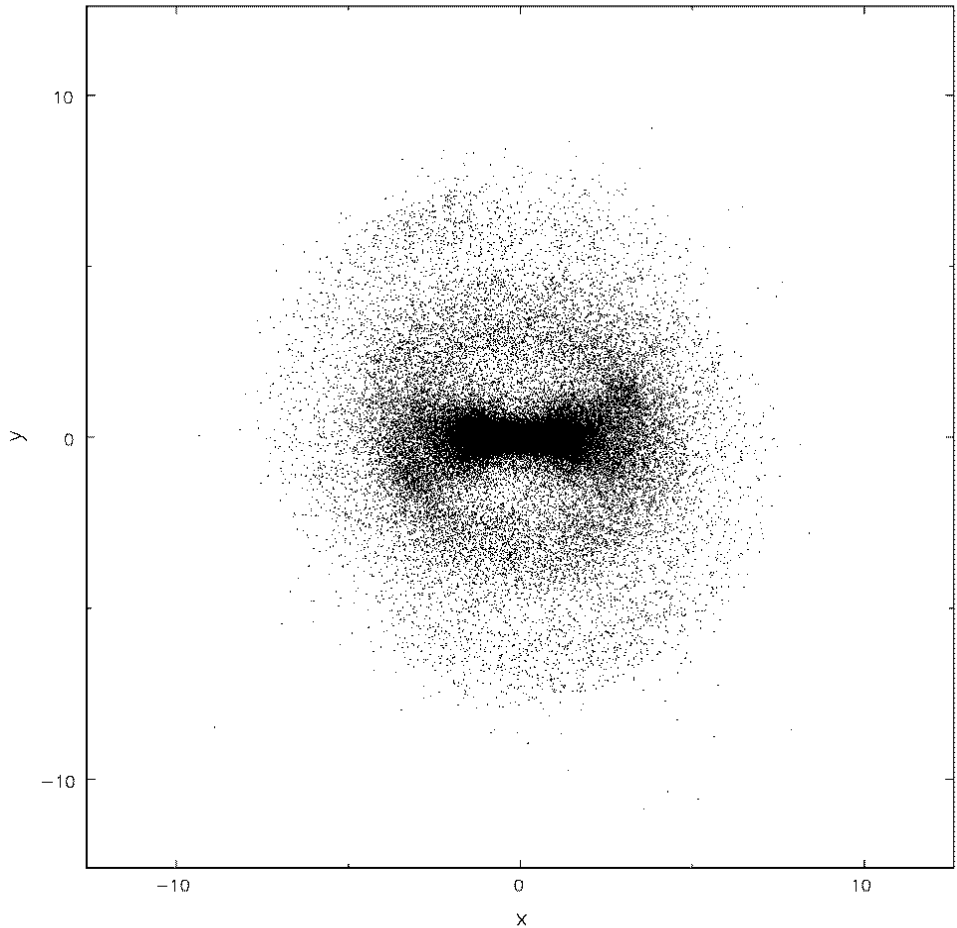,width=0.8\hsize,clip=}}
%\vspace{2in}
\caption{The disk particles in the canonical simulation at $t=1000$.  Note the prominent ring just outside the bar and the oval outer disk.  The box marks the full extent of the grid.}
\label{pntplt}
\end{figure}
\fi

\ifodd\style
\begin{figure}[t]
\centerline{\psfig{figure=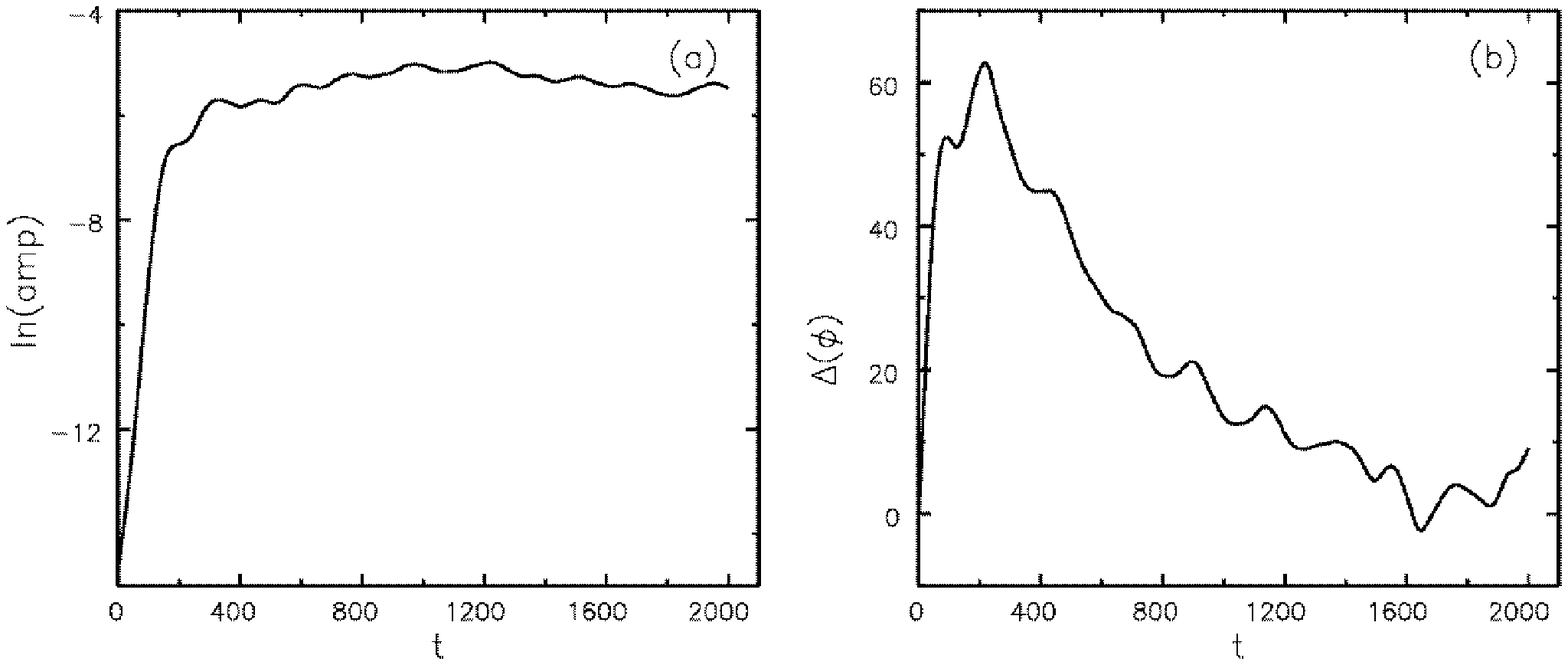,width=0.8\hsize,clip=}}
\caption{(a) The amplitude of an $m=2$ coefficient of the halo response in the canonical run. (b) The phase difference between the bar and the halo response.}
\label{fig:phaselag}
\end{figure}
\fi

\subsection{Canonical simulation}
\label{canonsim}

Our canonical simulation (run 4) is the most halo-dominated model reported in Paper I, which we describe in more detail here.  The initial rotation curve for this sub-maximal disk model (Figure \ref{4rotcurs}a) drops unrealistically beyond the disk edge because the halo density drops smoothly to zero at the grid edge.  The initial properties, numerical parameters and principal result are given in Tables \ref{tab:numpars} \& \ref{tab:allruns}.

The disk quickly forms a bar, as shown in Figure \ref{4abcd}(a).  Shortly after its formation, at $t \sim 150$, the bar buckled (Combes \& Sanders 1981; Raha \etal\ 1991) very mildly, causing it to thicken.  The continuing slow rise in $\aqm$ after this time results from gradual trapping of additional disk particles into the bar, often associated with spiral activity (Sellwood 1981); we describe this process as secondary bar growth.  Spiral activity gradually declines as the outer disk heats to $Q\gtsim 4$.  Figure \ref{cont10000} shows contour plots of the disk at several instances which clearly reveal the thickening of the disk, and the peanut shape of the bar.  Figure \ref{pntplt} shows the distribution of disk particles at $t=1000$; note the prominent inner ring and the oval outer disk.

Once the bar forms, $\om$ has a well-defined value (Figure \ref{4abcd}c) which drops rapidly at first, but reaches a constant level towards the simulation's end at $t=2000$.  Figure \ref{4abcd}(d) shows that $\len$ increases only mildly, whereas $\lag$ increases rapidly at first, later reaching a constant value, reflecting the behavior of $\om$ in Figure \ref{4abcd}(c).

Figure \ref{4abcd}(b) shows that the drop in $\om$ is associated with a substantial transfer of angular momentum from the disk to the halo.  The torque which produces this angular momentum exchange arises from dynamical friction on the bar as it moves through the halo.  The bar induces an $m=2$ response in the halo which develops very soon after the bar forms (Figure~\ref{fig:phaselag}a).  The response lags the position angle of the bar (Figure~\ref{fig:phaselag}b) at first, but gradually shifts into alignment with the bar as the torque drops.

It is interesting that the bar survived such strong braking (cf.\ Kormendy 1979), which reduced its pattern speed by a factor of $\sim 5$.  Most theoretical work, starting from Contopoulos (1980, see Sellwood \& Wilkinson 1993 for a review) has suggested that self-consistent bars should nearly fill their corotation circles; our simulation is a clear counter-example.  Its pronounced butterfly-shape when seen from above may be consistent with the prediction by Teuben \& Sanders (1985) that slow bars require a large fraction of ``box'' orbits.

The rearrangement of angular momentum altered the density distributions in both the disk and halo, causing the rotation curve to change to that shown in Figure \ref{4rotcurs}(b).  The central density of the disk rose significantly but the density profile of the halo barely changed, despite all the work done on it by the bar (Figure \ref{4abcd}b).  Such a small change underscores how difficult it is for any dynamical interaction with the disk to modify the halo density profile.

We emphasize that a more realistic flat rotation curve model would require a more massive and extended halo.  Not only might this increase dynamical friction, and slow the bar still more, but also the Lagrange point would lie further out in the disk, increasing the value of $\s$; this effect alone would increase $\lag$ by more than 30\% in this run at $t=2000$.

\subsection{Halo rotation}
\label{halorot}

All our models reported in Paper I, including our canonical model, have isotropic halos with no initial net angular momentum.  Here we turn our attention to the effects of halos with net rotation.  At first, we simply change the sign of $J_z$ for some halo particles according to the rule described in Appendix \ref{apphaloangmom}.  

\ifodd\style
\begin{figure}[t]
\centerline{\psfig{figure=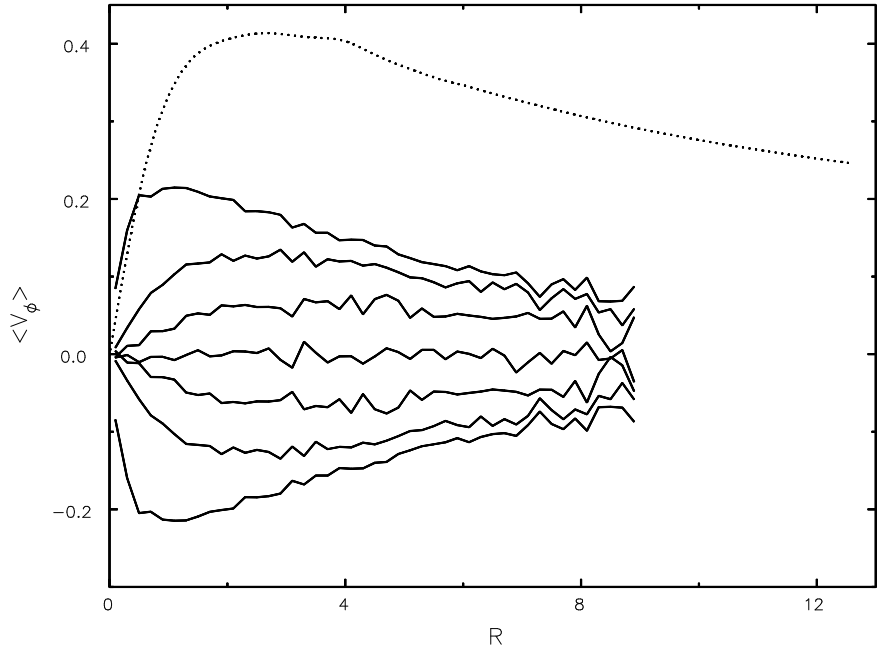,width=0.7\hsize,clip=}}
\caption{The mean rotation rates (full-drawn lines) of the halo particles in the simulations discussed in \S\ref{halorot}.  The circular speed in the mid plane is shown by the dotted curve.}
\label{fig:halorot}
\end{figure}
\fi

We ran a series of experiments, summarized in Table \ref{tab:allruns}, in which the halo angular momentum, ${\cal H} = L_{z,{\rm h}}/L_{z,\rm max}$, was $\pm 0.33, \ \pm 0.67$ and $\pm 0.98$, where $L_{z,\rm max}$ is the maximum achievable if the angular momentum of every particle is made positive.  (The corresponding values of the dimensionless spin parameter, $\lambda \equiv L |E|^{1/2}/(G M^{5/2})$ are 0.05, 0.11 and 0.16, respectively).  The mean rotation speeds of the halo particles are shown in Figure \ref{fig:halorot}.  Note that we did not continue all these simulations until the bar had finished slowing down.

Figure \ref{svslhalo} shows that the value of $\s$ reached by $t=1000$ correlates strongly with the angular momentum content of the halo.  As found previously by Athanassoula (1996), the bars in models with retrograde halos are more strongly braked, while those in prograde halos less so, in comparison with the non-rotating case.  Nevertheless, even when direct rotation in the halo is maximized, $\s = 1.7 \pm 0.4$ by the end of the run despite the fact that bar was weaker.  The value of $\aqm$ in the maximally rotating models settled at some 70\% of its value in the non-rotating simulation, suggesting that secondary bar growth, with a concomitant increase in friction, may be enhanced by angular momentum loss to the halo.

\ifodd\style
\begin{figure}[t]
\centerline{\psfig{figure=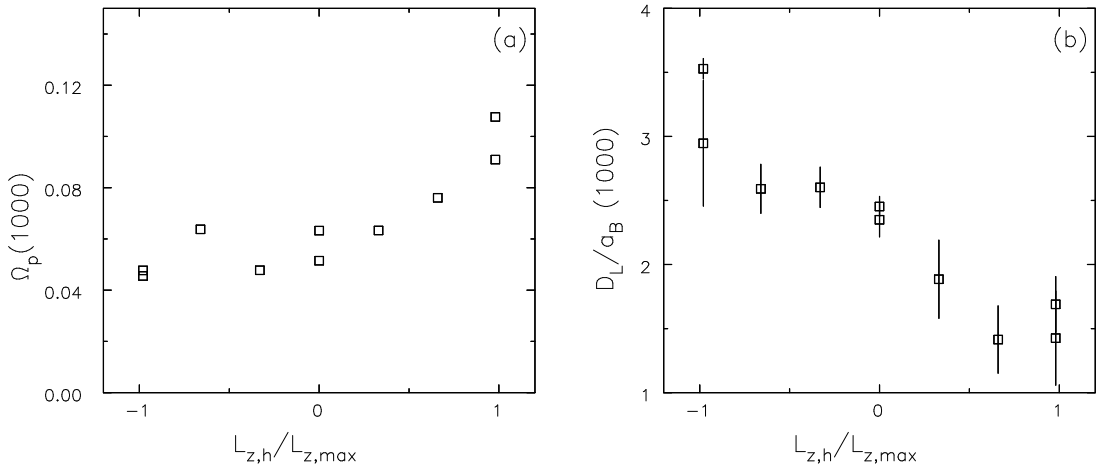,width=0.9\hsize,clip=}}
\caption{The variation of $\om$ and $\s$ with $L_{z,{\rm h}}/L_{z,\rm max}$ at $t=1000$ for all the simulations with isotropic halos that were set into rotation.  Note that there are two runs at each of $L_{z,{\rm h}}/L_{z,\rm max} = -0.98$, $0.0$ and $0.98$ giving some measure of the inherent scatter.}
\label{svslhalo}
\end{figure}
\fi

\subsection{Anisotropic halos}

We next present simulations with halos having somewhat more general DFs that yield anisotropic velocity distributions even in the absence of net rotation.  Inspired by Osipkov (1979) and Merritt (1985), we chose the form ${\cal F}(E + \beta J_z^2)$, where $\cal F$ has the same form as in equation (\ref{eqn:df}).  The halos are oblate and azimuthally biased when $\beta<0$ and prolate and radially biased when $\beta>0$.  Anisotropy changes the distribution of mass within the halo; we therefore adjusted the halo mass fraction, $f_{\rm h}$, and/or the halo truncation radius, $r_{\rm halo}$, to generate models with values of $\eta$ similar to that in our canonical simulation.

Some initial tests revealed that too pronounced an azimuthal bias ($\beta \leq -0.2$) led to strongly lop-sided halos which interfered considerably with the development of the bar.  These $m=1$ instabilities, which appear to be of the kind discussed by Sellwood \& Valluri (1997) for counter-rotating oblate spheroids, produced much larger asymmetries than those generally observed in real galaxies.  We therefore report only those simulations which did not become strongly lop-sided.  Disk-halo interactions via such $m=1$ modes are interesting in their own right and deserve a separate study.

The possible parameter space when the halos are allowed to be anisotropic is very large; we considered only two main models, an azimuthally biased model near the limit of $m=1$ stability, and a radially biased model (Table \ref{tab:allruns}).  Their rotation curves are shown in Figure \ref{anisorotcurvs}.  In both cases, we also tested fully-rotating versions of these models, which should have the weakest friction (\cf\ \S\ref{halorot}).

\ifodd\style
\begin{figure}[t]
\centerline{\psfig{figure=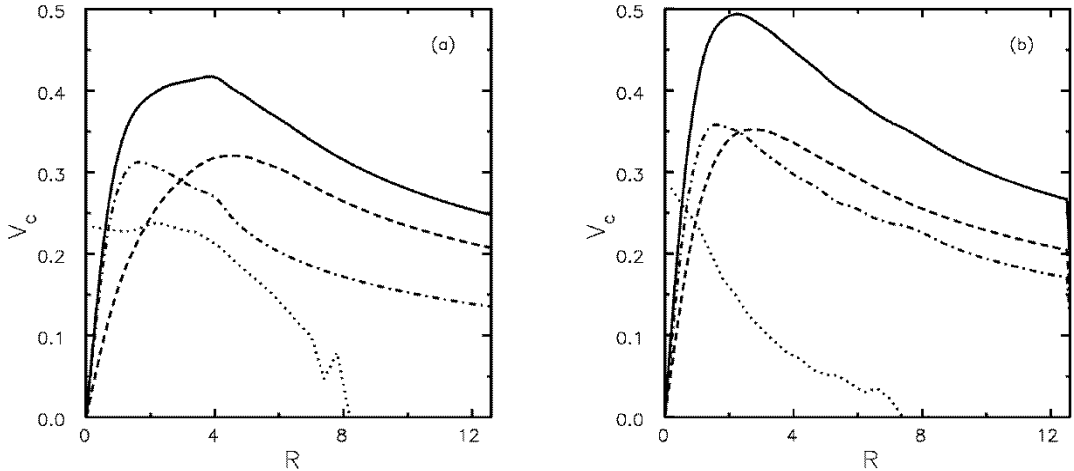,width=0.9\hsize}}
\caption{Rotation curve decomposition of the models having (a) the azimuthally biased/oblate halo and (b) the radially biased/prolate halo.  The solid line is the total rotation curve, the dot-dashed line is the disk contribution and the dashed line is the halo contribution.  The mean rotation speeds of the halo particles in the fully-rotating models are shown by the dotted lines.}
\label{anisorotcurvs}
\end{figure}
\fi

The bias towards large angular momenta introduced by setting $\beta<0$ tends to place more halo material at large radii at the expense of small radii and the inner rotation curve becomes dominated by the disk.  To compensate, we further decreased the radial extent of the azimuthally biased halos.  The oblate halo has an axis ratio $\simeq 0.7$ at $6\,R_{\rm d}$.  The bar that formed in a non-rotating halo was strongly braked, but friction is greatly reduced in the run with maximum halo rotation ($\lambda = 0.23$).

The halos generated by radially biased DFs have larger densities at smaller radii, all other things being equal, than do isotropic halos.  We therefore reduced the halo mass fraction in order to avoid a system which was too halo dominated.  The halo was prolate at large radii, having an axis ratio $\simeq 1.16$ at $4\,R_{\rm d}$, but it becomes oblate at $R_{\rm d}$ because of the disk's influence.  Since a trial run showed that secondary bar growth can be quite rapid in this model, we extended the disk to $R_t = 8\,R_{\rm d}$.  Once again, the bar is strongly braked in a simulation with no net halo rotation, but remains fast when the halo rotates maximally ($\lambda = 0.13$).

\subsection{Radially varying anisotropy and rotation}

\ifodd\style
\begin{figure}[t]
\centerline{\psfig{figure=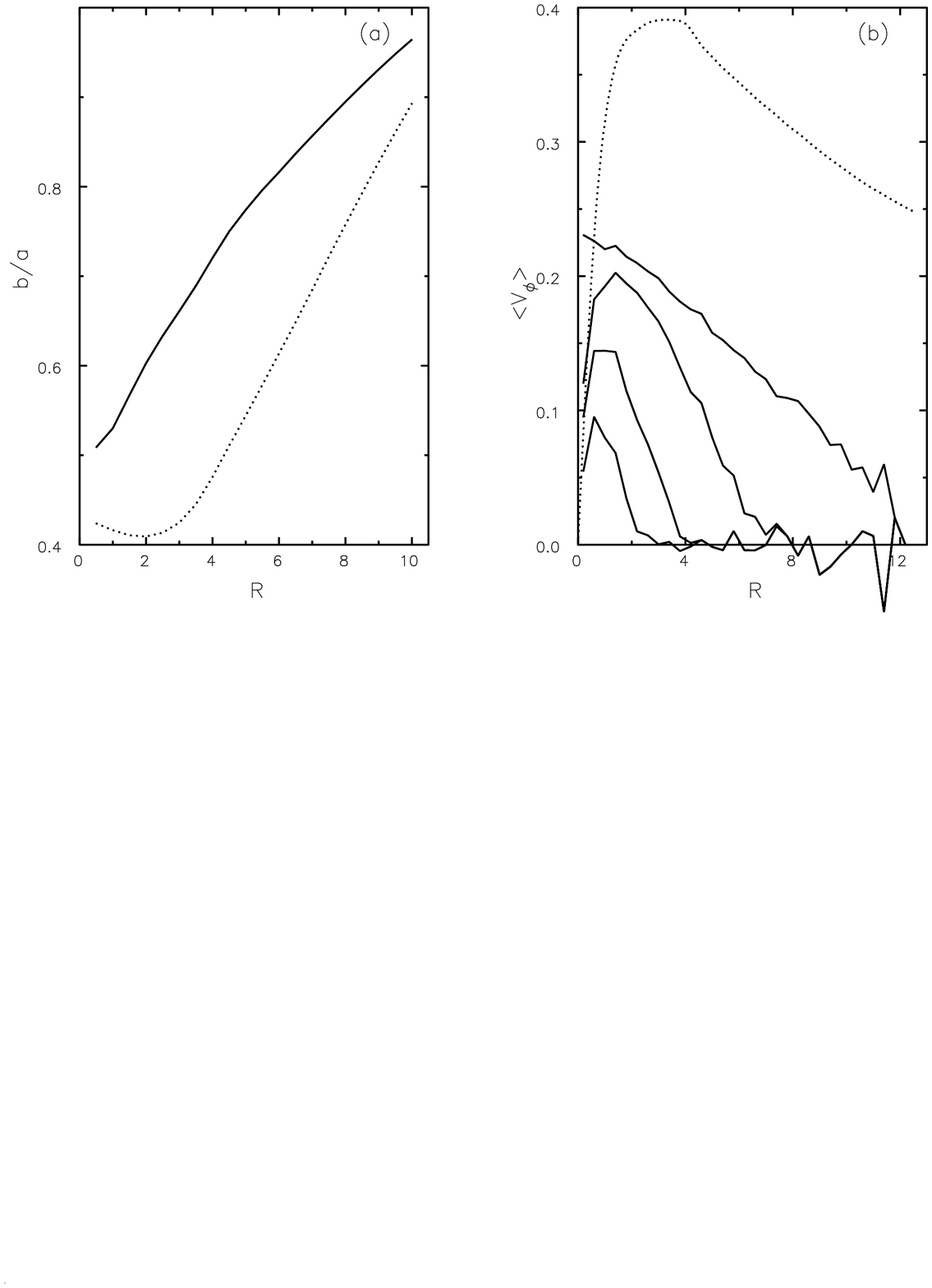,width=0.8\hsize,clip=}}
\caption{(a) The halo flattening for $\beta_0 = -0.2$ (full-drawn line) and for $\beta_0 = -0.5$ (dotted line).  These halos vary from nearly spherical at large radii to highly flattened in the inner parts.  (b) The initial mean orbital speeds of the halo particles (full-drawn curves) in four simulations of the $\beta_0 = -0.2$ case with rotating halos (see text).  The dotted curve shows the circular orbital speed in the disk.}
\label{run123aniso}
\end{figure}
\fi

Tremaine \& Ostriker (1999) suggest that the stringent limit on the halo central density that we obtained for isotropic halos (Paper I) could be relaxed if the halo had significant rotation in its inner parts only.  They proposed that the inner halo had itself been torqued up and flattened by interactions with the disk.

To test their hypothesis, we create halos with varying anisotropy.  We use polytrope-like distribution functions as before, ${\cal F}(E + \beta J_z^2)$, but now we let $\beta$ itself be a function of energy, $\beta(E)$:
\begin{equation}
\beta(E) = \beta_0 ( 1-\varepsilon^2)
\end{equation}
where $\varepsilon = (E-E_{\rm min})/(E_{\rm max}-E_{\rm min})$.  Here, $E_{\rm max}$ and $E_{\rm min}$ are the values of the potential at the grid edge in the disk plane and the center of the system respectively.  We set the free parameter $\beta_0 = -0.2$ in order to generate azimuthally biased, oblate halos, in line with the prediction of Tremaine \& Ostriker.  Figure \ref{run123aniso} shows the axis ratio of the halo as a function of radius, which varies from $\sim 0.5$ at the center to spherical at the edge.

We ran a set of three experiments (runs 137-139 listed in Table \ref{tab:allruns}) in which we introduced rotation by flipping retrograde halo particles with a probability that was a function of energy given by equation (\ref{eq:flipf}), which gives something close to maximal rotation in the halo to some limiting energy; beyond some spherical radius $r_{\rm rot}$ the halo is non-rotating (Figure \ref{run123aniso}b).  We varied this critical energy in this set of experiments.  Two other experiments (runs 123 \& 124) bracket them in terms of angular momentum content by having, respectively, no halo angular momentum and the maximum possible.

\ifodd\style
\begin{figure}[t]
\centerline{\psfig{figure=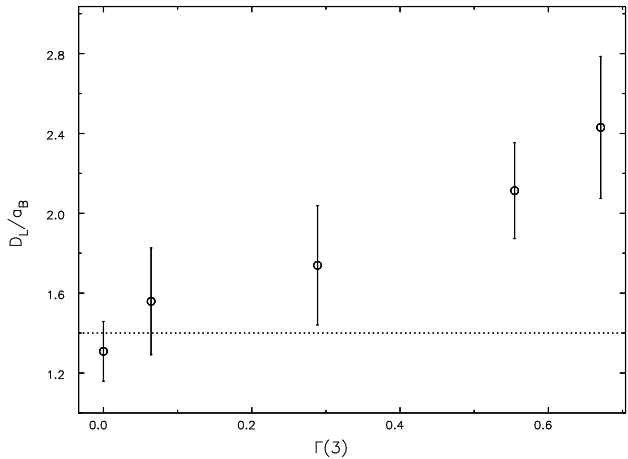,width=0.5\hsize,clip=}}
\caption{The variation of $\s$ with $\Gamma(3)$ for the $\beta_0=-0.2$ cases.  The largest value of $\Gamma$ is for a non-rotating halo and the dotted line marks $\s = 1.4$.  The values are for $t=1000$ or $t=1250$ -- the later time in simulations in which friction is still acting.}
\label{gmvss}
\end{figure}
\fi

Again, the results are given in Table \ref{tab:allruns}.  As the angular momentum content of the halo is increased, the bar slows down less, but even when $r_{\rm rot}=6$ (Figure \ref{run123aniso}b) the final value of $\s$ is still quite large.  The parameter $\s$ remains acceptably small only for the fully rotating case.

Mildly rotating halos in a similar set of experiments with a still more flattened halo ($\beta_0 = -0.5$), produced much weaker bars at first, because the bar buckled more violently; it seems that a different kind of coupling to the halo occurred in these cases.   As these weak bars were slowed to a lesser extent after a fixed amount of evolution, we continued one simulation to $t=3500$, and found that the bar recovered and substantial friction again developed leading to a slow bar ($\s = 2.26 \pm 0.20$).

The mean orbital speed of the halo particles (Figure \ref{run123aniso}b) already indicates that significant halo rotation out to quite a large radius is needed to avoid strong friction.  Figure \ref{gmvss} underscores just how much halo angular momentum this implies: we see that fast bars require $\Gamma(3) \ltsim 0.15$ which is $\sim 0.5$ less than the value for the non-rotating halo.  Thus the halo angular momentum inside $r=3$ has to be fully 50\% of that of the entire disk.

Tremaine \& Ostriker argue that strong halo rotation could be induced out to $\sim 3\;$kpc in a Hubble time, but we have shown here that the halo angular momentum requirements to avoid strong friction are considerably greater than their mechanism seems able to produce. 

% Apparently, as long as halo material from large radius can visit the region of the bar (\ie\ material at large energy having low angular momentum), the bar is still braked.

\section{Halo Density and Concentration}

\ifodd\style
\begin{figure}[t]
\centerline{\psfig{figure=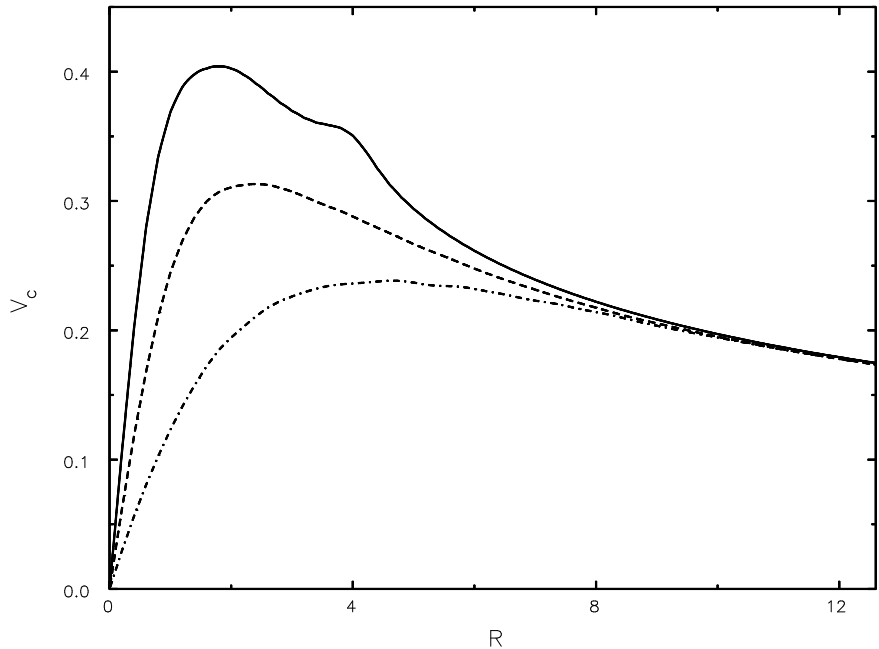,width=0.7\hsize,clip=}}
\caption{Contributions to the rotation curves from the $n = 1.6$ halo (dot-dashed), $n = 4$ halo (dashed), and the disk only (unbroken line).  The curves converge at large radii because these distributions all have equal total mass.}
\label{halorots}
\end{figure}
\fi

In this section we report experiments in which the halo mass and concentration are varied, still with the halos confined to small radii, as in \S3.  In \S\ref{maxdisk}, we discuss more realistic models with extended halos.  We have varied both the halo mass fraction and the polytrope index, $n$.  Increasing $n$ leads to more concentrated halos, resulting in a larger halo contribution to the inner rotation curve (Figure \ref{halorots}).  Table \ref{tab:allruns} lists the parameters of these simulations and gives our principal result.  Note that these simulations represent three series, with varying halo mass density at fixed $n$.

The evolution of the $n=3$ runs has already been presented in Paper I.  We find that decreasing the halo contribution to the rotation curve leads to a marked decrease in $\s$.  A fast bar (with a pattern speed that continued to decrease very slowly) survived in the simulation with the least massive halo of this series.

\ifodd\style
\begin{figure}[t]
\centerline{\psfig{figure=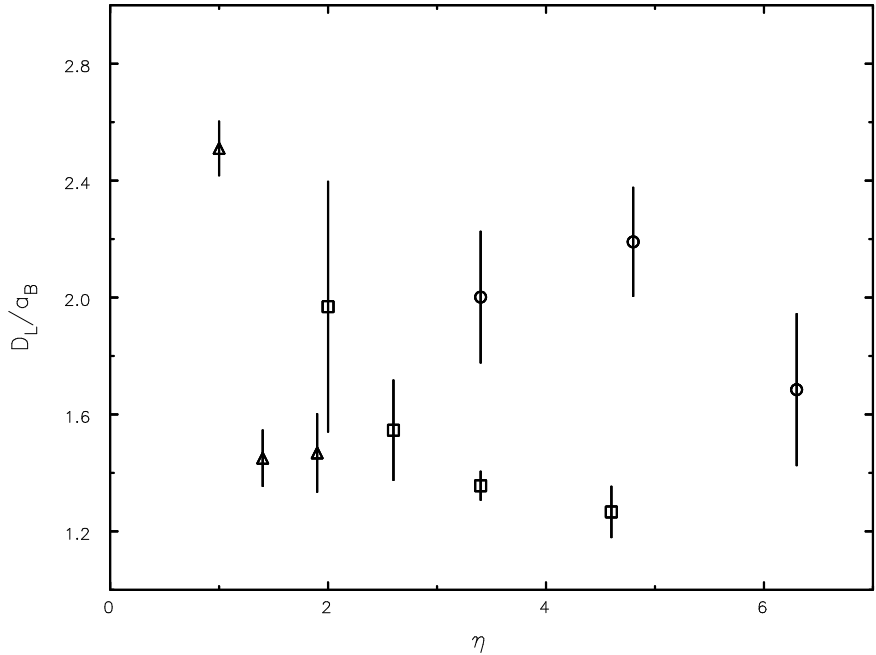,width=0.7\hsize,clip=}}
\caption{The variation of $\s$ with $\eta$ for different values of the polytrope index: triangles $n=4$, squares $n=3$, and circles $n=1.6$.  We plot averages of three values of $\s$ near the end of the simulations when friction is low, but not always zero.}
\label{svseta}
\end{figure}
\fi

Figure \ref{svseta} summarizes the values of $\s$ obtained from models with different polytrope indices.  Although there is no trend when all the points are taken together, within each series of runs at fixed $n$, bars with larger $\eta$ end up faster.  Note that it would be incorrect to conclude from this Figure that less braking occurs for a given $\eta$ as the central density increases.  This apparent trend results from two different effects:  First, increasing $n$ for fixed halo mass leads to greater halo concentration, depleting halo material at larger radii in our halos (which we confined to a small volume), thereby reducing friction somewhat.  At the same time, a more sharply peaked halo rotation curve leads to a smaller $\eta$ at fixed $M_{\rm h}$, and also makes the rotation curve drop more steeply, reducing $\lag$.

\section{Models with more Extensive Halos}
\label{maxdisk}

We next describe two models in which the halo extended to $r=25\,R_d$, which is large enough to achieve a flat rotation curve with moderate to low central densities in the halo.  We had to increase the grid to $257^3$ -- an eight-fold increase over that used in most of the above experiments, making these runs much more expensive.  A large parameter space search, such as that described in the previous two sections, would be prohibitively expensive with this larger grid.

We already reported a maximum disk model with an extensive halo in Paper I whose evolution was computed using a cylindrical polar grid.  One of the two models presented in this section closely resembles it, but has a different disk type and is run on a Cartesian grid.  The other model discussed here is a ``control'' experiment with a similar rotation curve but with a somewhat denser halo.

\ifodd\style
\begin{figure}[t]
\centerline{\psfig{figure=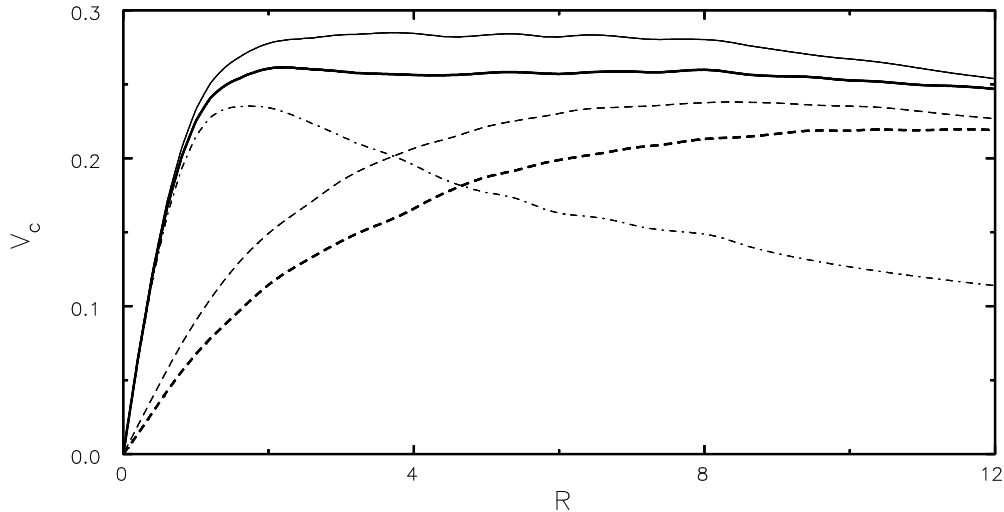,width=0.8\hsize,clip=}}
\caption{The unbroken curves show the circular speeds of the maximum disk (bold) and the control experiments.  The dot-dashed line shows the disk contribution (common to both experiments) and the dashed lines are the halo contributions.  Note that both rotation curves are almost flat out to $8 R_d$.}
\label{runs68141rotcurvs}
\end{figure}
\fi

The disk, which was truncated at $8\,R_d$, accounts for 17\% of the total mass.  The polytrope index $n=2$ for the maximum disk model, whereas $n=3$ for the control run, this being the only difference between the two simulations.  The resulting rotation curves, shown in Figure \ref{runs68141rotcurvs}, are quite flat out to the disk truncation radius.   Both have a substantial disk contribution at small radii: $\eta = 7.0$ in the maximum disk case and $\eta = 3.8$ in the control experiment.

The bars grew very rapidly and did not buckle much, reaching an amplitude some $\sim 5$\% lower than in our canonical run.  Most of the angular momentum lost by the inner disk after the bar forms ends up in the halo for both runs, but the outer disk continues to accept some of the bar's angular momentum.

\ifodd\style
\begin{figure}[t]
\centerline{\psfig{figure=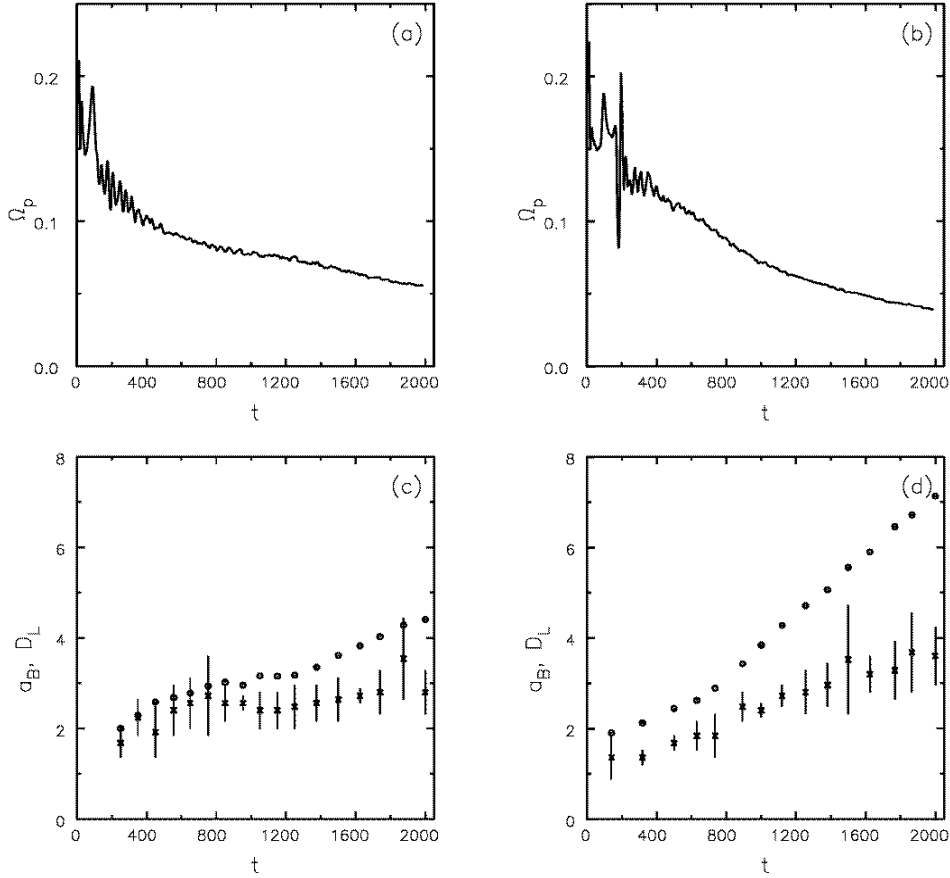,width=0.8\hsize,clip=}}
\caption{The upper panels show the evolution of $\om$ for the maximum disk and the control simulations while $\lag$ (circles) and $\len$ (crosses) are shown the lower panels.  The maximum disk is shown on the left (a \& c) and the control run on the right (b \& d).}
\label{runs68141speeds}
\end{figure}
\fi

Figure \ref{runs68141speeds} shows the evolution of $\om$, $\len$ and $\lag$ for both simulations.  The bar in the control simulation is slow already by $t \sim 1500$ and $\s = 1.98 \pm 0.35$ by the end.  The bar in the maximum disk simulation, on the other hand, is acceptably fast at the end of the run, with $\s = 1.57 \pm 0.27$, but only barely so and it is continuing to slow.  We have therefore identified a region of parameter space where fast bars can survive for more than $\sim 30$ orbital periods at $R=1.4$ (just outside the half-mass radius of the disk).

\section{Synthesis}

We now seek some way to synthesize all these numerical results.  We first show that the frictional torque acting on the bar from the halo behaves in some sense as might be expected from standard dynamical friction.  However, the total angular momentum loss which occurs before the halo response locks into phase with the bar can be determined only numerically; we attempt to relate the final pattern speed of the bar to the ability of the halo to accept angular momentum.

\subsection{The Chandrasekhar formula}

\ifodd\style
Chandrasekhar's (1943) demonstration of a secular drag force on a massive perturber moving in a straight line through a uniform, infinite background sea of low mass particles differs in many significant ways from the rotation of a bar through an inhomogeneous halo.  It is now part of received wisdom that his formula works better for a perturber moving through an inhomogeneous system, using local values of the density \etc, than we have any right to expect (Binney \& Tremaine 1987, \S7.1).  We here show that the frictional torque also behaves very roughly in accordance with his formula for a rotating bar -- at least over the period after the bar has formed and settled and before the halo response becomes aligned with the bar.

\begin{figure}[t]
\centerline{\psfig{figure=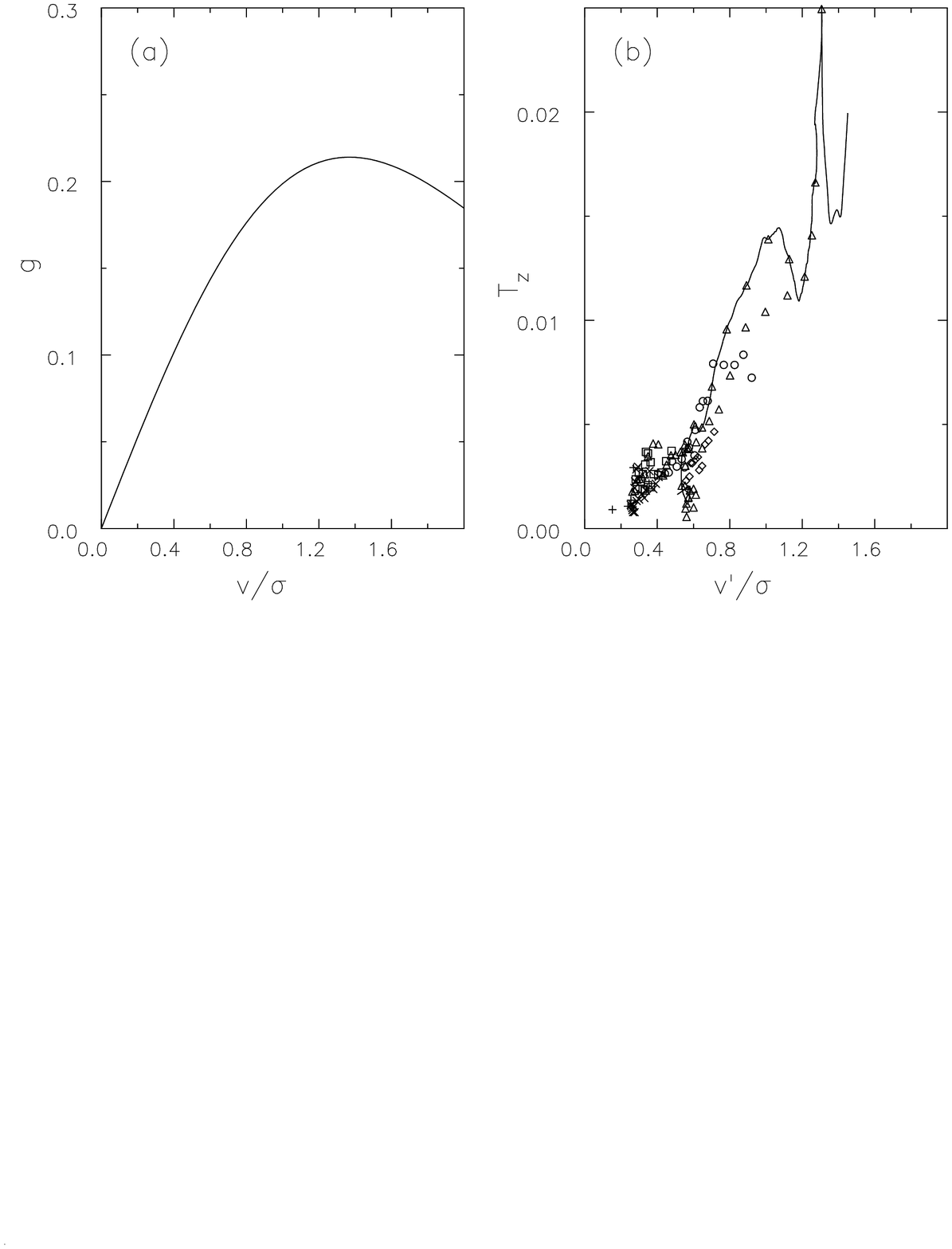,width=0.9\hsize,clip=,angle=0}}
\caption{(a) The function $g$ defined in equation (\ref{eq:thchandra}).  (b) The scaled halo torque $T_z$ (equation \ref{eq:torque}) measured from many simulations all having the same disk/halo mass.  See text for an explanation of the line.  The symbols are as follows: canonical model (squares), hot disk models (circles), retrograde halos (triangles), prograde halos (pluses), other polytrope indices (crosses), rotating halos with radially varying anisotropy (diamonds).}
\label{fig:chandra}
\end{figure}
\fi

In the limit when the perturber's mass $M$ is much larger than the masses of the background particles, Chandrasekhar's formula for the frictional force is
\begin{equation}
M{d \bv \over dt} = -\hat{\bv} {4\pi\ln\Lambda G^2 \rho M^2 \over \sigma^2} g\left({v \over \sigma} \right),
\label{eq:thchandra}
\end{equation}
where $\bv$ is the velocity of the perturber, $\hat{\bv}$ is a unit vector in the same direction and $\rho$ and $\sigma$ are respectively the density and velocity dispersion of the background.  The $v^2$ factor in the denominator of formula (7-18) of Binney \& Tremaine has here been replaced by $\sigma^2$ in order to subsume all the velocity dependent factors into the dimensionless function $g$, which is shown in the Figure \ref{fig:chandra}(a) for the case when the velocity distribution of the background particles is Gaussian.  It can be seen that friction increases as the speed of the perturber rises, reaching a maximum when $v \simeq 1.37\sigma$ and then decays monotonically for all higher speeds.

The rate of gain of angular momentum of the halo in our simulations is clearly the frictional torque $\tau_z$ on it arising from the bar.  Note that this measurement is independent of bar's reaction to the loss of angular momentum, and therefore does not involve, for example, any estimate of its effective moment of inertia.  The halo torque could be interpreted as the frictional force times some characteristic lever arm of length $R$.  We therefore plot in Figure \ref{fig:chandra}(b) the quantity
\begin{equation}
T_z = {\langle\tau_z\rangle\sigma^2 \over R \langle {\aqm} \rangle^2 \rho}
\label{eq:torque}
\end{equation}
against the speed of the bar perturbation through the background halo at that radius, $v^\prime = R\om - \langle v \rangle$, normalized by the halo velocity dispersion.  The running averages in this expression are over 50 time units, and we include $\aqm$ to take account of the variations in bar mass in these equal mass disks.  The scaling of the ordinate is arbitrary, therefore.  We adopt $R=3\peak$ (\ie\ three times the radius where the $m=2$ Fourier component peaks); other values of $R$ show the same general trend, but we found that this choice minimized the scatter.  We adopt local values for $\rho$ and $\sigma$ (averaged over the range $0 < R < 3\peak$), and we use $\sigma^2 = \sigma_R^2 + \sigma_\phi^2 + \sigma_z^2$.

We plot a curve from one simulation in Figure \ref{fig:chandra}(b) and a number of points from other simulations.  The curve shows the entire evolution of $T_z$, from the moment the bar first forms to the end when friction is very low, for a maximally retrograde model (run 20).  The time evolution along this line is from right to left and can be described as follows: (1) the initial spike occurs as the halo response develops soon after the bar forms, (2) the curve then dips as the bar buckles, (3) after which the line follows roughly the trend indicated by the points as the bar is braked steadily, and (4) it drops down to low values as the halo response locks into alignment with the bar.  The fluctuations in $T_z$ in this run are fairly typical; they are larger in some models and less in others.

We should not expect Chandrasekhar-style friction in any part of this evolution except for period (3) after a quasi-steady halo response is established, and before orbit trapping becomes significant.  Thus we have tried to include points in this plot from other runs during the steady friction period, although we obviously failed for the cluster of points near $v^\prime/\sigma =0.6$ and $T_z \gtsim 0$.

Although there is considerable scatter in these measurements, the general rise over the range of the abscissae is similar to the theoretical curve in Figure \ref{fig:chandra}(a), although our data show only a rising trend.  We are encouraged that a trend shows through despite the crude approximations we adopt: we identify a single radius whereas the entire halo probably contributes, not all our bars have identically the same density profile, our polytrope halos do not have a precisely Gaussian velocity distribution, \etc\ ~While perhaps not entirely convincing, this Figure does suggest (i) that the torque is indeed from the usual dynamical friction, (ii) that most of the drag seems to arise from the region just beyond the end of the bar, and (iii) that the characteristic speed is generally less than the halo dispersion.

It should be noted that the velocity dependence in Figure \ref{fig:chandra}(b) is the opposite of that predicted by Weinberg (1985), who suggested that friction would increase if the halo was made to rotate in a prograde sense; one interpretation of Weinberg's prediction is that in his case most of the friction arises from a perturber speed $v \gtsim 1.4\sigma$.  The fact that we find the opposite behavior may result from his assumption of an infinite isothermal sphere for the halo, whereas our halos have a very limited radial extent.

\subsection{Constraints from $\Gamma$}

\ifodd\style
\begin{figure}[t]
\centerline{\psfig{figure=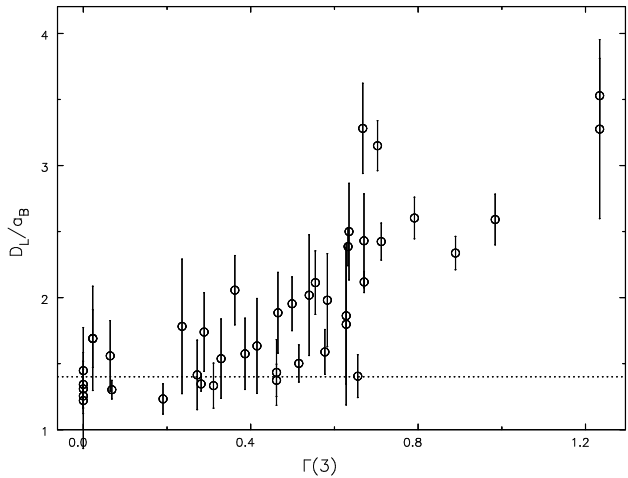,width=0.5\hsize,clip=}}
\caption{The distribution of 43 simulations in the $\Gamma(3)$, $\s$ plane.}
\label{gammpar}
\end{figure}
\fi

In our simulations with strongly confined halos, friction seems to drop to zero before the bar is brought to rest relative to the streaming speed of the halo particles.  The torque vanishes when the induced bi-symmetric distortion in the halo locks into alignment with the bar.  This locking effect appears to be the result of non-linear trapping of orbits, a process described as ``dynamical feedback'' by Tremaine \& Weinberg (1984b).  Note that this locking phenomenon did not occur in our simulations with more extended halos, where braking persisted for as long as we ran them (see Figure \ref{runs68141speeds}).

We have searched for a predictor of the final bar pattern speed, but have not found anything simple -- perhaps because none exists.  The best we have come up with is the parameter $\Gamma$ introduced in \S\ref{sec:params}.  After some experimentation, we found $\Gamma(3)$, evaluated using the initial values of $J_{z,h}$ and $J_{z,d}$, to be the most useful.  Figure \ref{gammpar} summarizes the last measured values of $\s$ from an ensemble of 43 different simulations plotted against $\Gamma(3)$.  These simulations include various different polytrope indices, different anisotropies, different halo rotations (and different distributions of halo angular momentum), different Toomre $Q$s, different disk thicknesses and different halo masses.  Not all of these simulations have been evolved to a steady state; we generally stopped the simulation once we found the bar to be slow (which may account for some of the scatter in the distributions).  

We draw the following conclusions from the rising trend in this Figure: (1) Bars can generally remain fast when $\Gamma(3) < 0.4$, although some are significantly braked.  (2) When $\Gamma(3) \gtsim 0.4$, bars generally end up slow.  Strong braking can be avoided when $\Gamma(3) \gtsim 0.4$ when either the model has a rapidly dropping rotation curve, or a flattened halo with most of its mass outside the bar region.\footnote{The positions of models with radially varying anisotropy in this plot is more than usually sensitive to the value of $r_0$ used in computing $\Gamma$.}

Our parameter $\Gamma$ is the best we have been able to find to predict the value of $\s$.  A range of values of $\Gamma$ can be determined for any rotation curve fit, but, being dark, no value of $\Gamma$ can be pinned on a halo.  Constraints on the actual angular momentum content of dark halos might be possible from studies of the faint luminous halos that may trace the rotation of the dark halo (being subject to similar dynamics).  The Tremaine-Ostriker hypothesis calls for low $\Gamma$ due to high rotation in the halo.  We favor low $\Gamma$ due to low halo mass fraction (we have argued that the Tremaine-Ostriker hypothesis may need rather unlikely levels of angular momentum in the inner parts of halos).  The reality may be somewhere in between these two limits.

\section{Discussion}

\subsection{Comparison With Real Bars: NGC 936}
\label{sec:comparison}

We need to show that our $N$-body bars are similar in strength to bars in real galaxies.  A photometric comparison would not be conclusive because the strongly non-axisymmetric light distribution in a galaxy may not reflect the true distribution of mass.  We have therefore attempted a kinematic comparison using data from \n936, a well-studied SB0 galaxy having sufficient published data on the stellar velocity field for our purposes.  Its other advantages are that it appears to be relatively dust-free and is known to have a fast bar (Kent 1987; Merrifield \& Kuijken 1995).  The galaxy has a dense bulge (Kent \& Glaudell 1989), however, unlike in our simulations.

We adopt Kormendy's (1983) estimates of the projection geometry, the bar position angle, the de-projected bar semi-major axis and use his velocity measurements at five different slit position angles.  We scale our models by setting $\len = 50^{\prime\prime}$ and rotate and project them as we view \n936.  We then compute the mean projected line-of-sight velocity ($V_{\rm los}$) of the particles in the model, averaging the approaching and leading sides to maximize the number of particles in our pseudo-slits, and determine the velocity scaling by minimizing $\chi^2$ between the observations and the model.  When making this comparison, we use data in the circular annulus (in the galaxy plane) $0.6 \leq R / \len \leq 1.2$ to exclude the region dominated by the bulge and the disk well outside the bar.

Our canonical model at early times compares very well with \n936.  We find reduced $\chi^2 = 0.70$ (for 22 degrees of freedom) at $t=250$, and a visual comparison of $V_{\rm los}$ shows that the variations with position angle track those in the galaxy very well, as they must do for this good a fit.  (For comparison, we obtain a reduced $\chi^2 = 1.96$ if we erase the bar from our model by randomizing the azimuthal positions of the particles.)  At later times, after the bar has been slowed by a factor $\sim 5$, we find reduced $\chi^2 = 5.94$, again showing that \n936\ is grossly inconsistent with a slow bar.

The maximum disk model is not quite as impressively similar to \n936: reduced $\chi^2 = 1.20$ at $t=350$ which is still acceptable, but reduced $\chi^2 = 1.54$ at $t=1150$, which is marginally so.  The fits to ``control'' run are again worse: reduced $\chi^2 = 1.61$ at $t=318$, when the bar is still fast, and reduced $\chi^2 = 1.75$ at $t=2000$ when the bar has slowed.  In both cases, the absence of a massive central spheroidal bulge may be partially responsible for the poorer fits.

Thus the bars in our simulations are quite similar to that in \n936 when they first form, but are clearly inconsistent when they have been strongly braked.

\subsection{Neglected Physics}

Our simulations are of the stellar and DM components of a barred galaxy and do not include any gas.  It is well known that gas flows in barred potentials produce large-scale shocks offset from the bar major axis (\eg\ Athanassoula 1992).  The asymmetric gas distribution loses angular momentum to the bar and gas flows towards the center.  The small gas fraction in most galaxies, together with the relatively short lever arm, mean that the angular momentum given to the bar by gas could not possibly compete with that lost to the halo through dynamical friction -- \eg\ friction removed $\sim 40$\% of the disk's angular momentum in our canonical simulation (Figure \ref{4abcd}).  Furthermore, gas-poor SB0 galaxies, such as \n936 and NGC~4596, have fast bars.

Gas inflow has a second effect, however: it deepens the gravitational potential at the bar center causing the bar to speed up slightly.  This can be a small effect at most, since excessive inflow will destroy the bar.  The total mass accumulated in the center cannot exceed 5\% of the disk mass (Norman \etal\ 1996), and is probably much less; even in this extreme case, the increase in bar pattern speed was only some 40\% (Sellwood \& Debattista 1996, Figure 6).

A bulge component inside the inner Lindblad resonance might act as a source of angular momentum for the bar.  Bulges can be quite massive (\eg\ \n936, Kent \& Glaudell 1989) and are often in rapid rotation (\eg\ Kormendy \& Illingworth 1982), but their small radial extent limits their angular momentum content.  It is possible to think of the inner parts of the rapidly rotating halo in some of our simulations as representing a bulge.  The bar is still strongly braked in such cases (\S\ref{halorot}), suggesting that not even rapidly rotating bulges can prevent bar slow-down in sub-maximal disks.  Weinberg (1985) suggested that the inner disk could also be a source of angular momentum for the bar, but we have found that bars always {\it lose\/} angular momentum to the disk, not the other way round.  

We have not included the effects of late gas infall onto the disks which would enhance secondary bar growth (\eg\ Sellwood \& Moore 1999).  Irrespective of the rate of bar growth by this mechanism, it inevitably leads to increased friction, making it still more difficult for the value of $\s$ to decrease.

In addition to massive gas inflow, bars can be destroyed by satellite impacts.  Our simulations do not take either process into account.  If these processes are to account for the absence of slow bars, they would have to act efficiently and quickly, and new bars would have to form again to maintain the observed high fraction of barred galaxies.  The rapid formation of a new bar is difficult to arrange, since bar destruction leaves the disk dynamically hot, and the inflow destruction mechanism gives rise to a more steeply rising inner rotation curve.  Both factors limit the disk's ability to form new bars, by making it less responsive and by cutting the feed-back loop to the swing amplifier (Toomre 1981).  The revival of a bar after one has dissolved requires the accretion of so much dynamically cool material (Sellwood \& Moore 1999) that it is unlikely to occur more than once in a galaxy.

While the bars in our simulations are comparably strong to that in a real galaxy (\S\ref{sec:comparison}), a systematic difference with early-type galaxies is the absence of a dense bulge in our simulations.  It is possible that simulations with dense bulges behave differently, but it seems unlikely that they would.  Orbit studies (Athanassoula 1992) and simulations (Sellwood 1989; Sellwood \& Moore 1999) reveal that bars in galaxies with dense centers are supported by the same orbit family and behave similarly to those in which the center is more uniform.

\subsection{Scaling to NGC 3198}

\ifodd\style
\begin{figure}[t]
\centerline{\psfig{figure=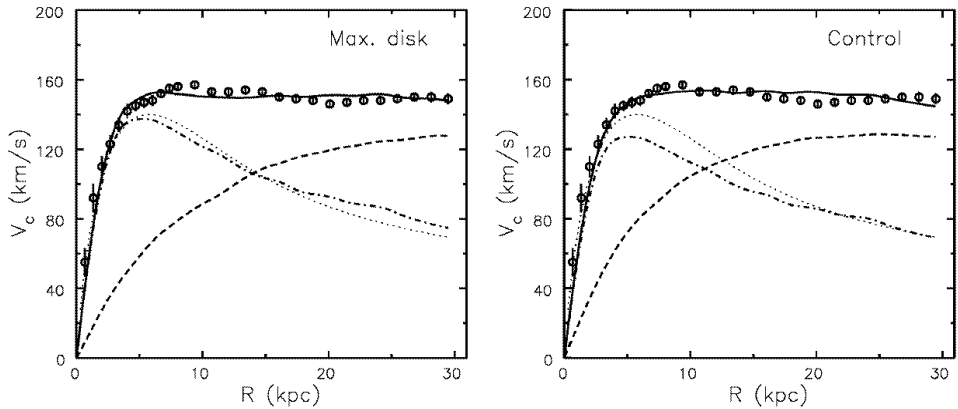,width=0.8\hsize,clip=}}
\caption{The ``maximum disk'' and ``control'' simulations scaled to NGC 3198.  The data for NGC 3198, shown in circles with error bars, are from van Albada \etal\ (1985).  The solid lines show the full rotation curves, the dash-dotted lines the disk contribution and the dashed lines the halo contribution.  The dotted line is the maximum disk as defined by van Albada \etal}
\label{ngc3198}
\end{figure}
\fi

Before discussing the implications of our results for real galaxies, we need to determine how they should be scaled.  The {\it de facto\/} standard galaxy in the dark matter halo debate is NGC 3198; even though it is {\it not\/} a strongly barred galaxy, we nevertheless scale our models to the data of van Albada \etal\ (1985) for this system.

We use our two extensive halo systems from \S\ref{maxdisk} for this comparison.  Scaling to the observed rotation curve determines the length and the velocity scales.  We first adopted $R_d = 3.0$ kpc for both simulations, which differs from the exponential scale of 2.68 kpc that fits NGC 3198 well (van Albada \etal\ 1985) since our models have Kuz'min-Toomre disks.  We then adjusted the velocity scale to minimize the residuals between the data and our model rotation curves, finding $(GM/R_{\rm d})^{1/2} = 584$ \& 540~km~s$^{-1}$ for the maximum disk and control runs, respectively.  The resulting scaled rotation curves are shown in Figure \ref{ngc3198}, which also shows the maximum-disk fit of van Albada \etal\ \ As usual, both models fit NGC 3198 very well, which is another instance of the rotation curve degeneracy.  Note that both systems are quite disk dominated, and that the maximum disk of van Albada \etal\ is perhaps even more disk dominated than our ``maximum disk'' simulation.

Choosing length and velocity scales determines the time unit.  With these adopted values, the duration of many of our experiments, $2\,000$ dynamical times, is equivalent to $\sim 10$ Gyr.

\section{Conclusions}

\subsection{Summary of Principal Results}

We have shown that the severe braking of the bar by dense isotropic halos reported in Paper I also occurs for other non-rotating, or backwards rotating, halos of similar density, whatever the shape of the halo velocity ellipsoid.  In all such cases, the bar in the disk slows unacceptably in a few rotations.  The bars in all our experiments with strongly prograde rotation in the halo were not braked as severely; the halo spins strongly in those cases for which friction ceased before $\s$ rose above 1.4.

The existence of strong friction is in agreement with the theoretical prediction by Weinberg (1985), with earlier fully self-consistent simulations by Sellwood (1980), using a coarse grid, and by Athanassoula (1996) using a different $N$-body method, and other work (\eg\ Hernquist \& Weinberg 1992).  We also find that all our bars slow down as they lose angular momentum -- a non-trivial result since bars are not rigid objects and could conceivably spin up (\eg\ as a binary star) as angular momentum is lost.  Remarkably, the bars in some of our experiments survived a truly drastic slow-down -- more than a factor of five in many cases -- providing further evidence that bars are in fact dynamically very robust objects (Miller \& Smith 1979; Sellwood \& Wilkinson 1993).  We find no evidence to support the idea (Kormendy 1979) that strong braking of a bar might cause it to dissolve.

In some of our simulations, the bar did a great deal of work on the halo -- \eg\ in the canonical run the halo gained 40\% of the angular momentum of the disk.  Nevertheless, the change in the halo density profile was quite minor (Figure \ref{4rotcurs}b).  This example emphasizes that it is extremely difficult to change the density profile of a halo using interactions with the disk.

Friction is generally reduced by lowering the density of the halo, and bars in maximum disks are able to remain fast (though only barely so) for large numbers of dynamical times, even in extensive, non-rotating halos, as reported in Paper I.  The bar in our maximum disk simulation with an extensive halo is continuing to slow down even after 2000 dynamical times (Figure \ref{runs68141speeds}a), suggesting that $\s$ might continue to rise.  When scaled to NGC 3198, this continued evolution is too slow to matter, since we have followed it for 10$\,$Gyr.  But dynamical times are shorter in more luminous galaxies and our computed evolution lasts the equivalent of 5$\,$Gyr in a galaxy with $V_{\rm max} \simeq 300\,$km~s$^{-1}$.

Friction is caused by a lag between the bar and an $m=2$ distortion in the halo (Figure \ref{fig:phaselag}).  None of our halos is rotating sufficiently rapidly to be itself unstable to bar-forming modes (\eg\ Sellwood \& Valluri 1997), so the halo distortions are responses forced by the rotating bar in the disk, as is usually the case in dynamical friction.  It is encouraging that we have been able to find some suggestion of the expected velocity dependence of the frictional force in our very crude analysis (Figure \ref{fig:chandra}), which suggests that the qualitative effect of halo rotation is predictable.

Braking ceases once the forced response in the halo rotates in alignment with the bar in the disk.  The gradual trapping of halo orbits into the driven non-axisymmetric potential is itself one of the principal sources of dynamical friction.  It seems reasonable that trapping of halo orbits should involve less angular momentum loss for the bars in halos with an excess of particles with $J_z>0$, as we have found.  Note that we have observed the locking of the halo response into alignment with the bar only in models with strongly confined halos.

\subsection{Implications for barred galaxies}

As noted in the introduction, the rather small number of actual measurements of real barred galaxies all lie in the range $1\leq\s\leq1.4$; the existence and locations of dust lanes in bars is indirect evidence that these low values pertain more generally.  Thus our results require either that most strong bars are (1) too young to have been slowed significantly, (2) exist in strongly rotating halos, or (3) are not embedded in dense halos.  We review each of these in turn.

If disks are not maximal, and halos do not rotate strongly, then bars must indeed be young objects to have remained fast today.  Since the rate of bar slow-down scales with the halo density, the larger the required density, the younger they must be to avoid any slow cases.  There is a suggestion that bars were rare in the early universe (Abraham \etal\ 1999), but they have certainly existed, probably in about their present numbers, since a redshift of one half.  Their ages are therefore $\gtsim 4\,$Gyr, or $\gtsim 800$ dynamical times when our simulations are scaled to NGC 3198, which is plenty long enough for friction to have slowed the bars significantly, although perhaps not completely.

Bars in galaxies which are significantly sub-maximal can remain fast for cosmologically interesting times only if the halo is anisotropic and rotates strongly throughout most of the disk region.  The required halo angular momentum is very large, however.  If all halo angular momentum arises from tidal torques in the early universe, the required value of $\lambda$ would be reached only rarely (Barnes \& Efstathiou 1987; Steinmetz \& Bartelmann 1995).  Since $\gtsim 50\%$ of all HSB disks containing strong bars (Eskridge \etal\ 2000), the vast majority cannot have $\lambda$ large enough to avoid bar slow down.

Alternatively, one could imagine the inner halo to have been torqued up by some means.  Tremaine \& Ostriker (1999) suggest two ways to transfer angular momentum from the disk to the inner halo for precisely this purpose.  We have found that if the halo has moderate central density, then it must have a high degree of rotation -- fully half that of the disk, out to well beyond the bar's end.  If the nearest strongly barred galaxy, our own Milky Way, has a sub-maximal disk, we require substantial rotation in the halo within the Solar radius for the bar to have avoided strong braking.  Most torquing mechanisms would act equally on both the dark halo and any associated stars.  Thus the absence of significant rotation in, for example, the metal-weak globular cluster system of the Milky Way (Harris 2000) also suggests that the inner halo lacks the required angular momentum.

We therefore conclude that bars in real galaxies remain fast because disks are maximal.  Weiner \etal\ (2000) reach a similar conclusion on quite different grounds in the case of the barred galaxy NGC 4123.

\subsection{General Implications}

\ifodd\style
\begin{figure}[p]
\centerline{\psfig{figure=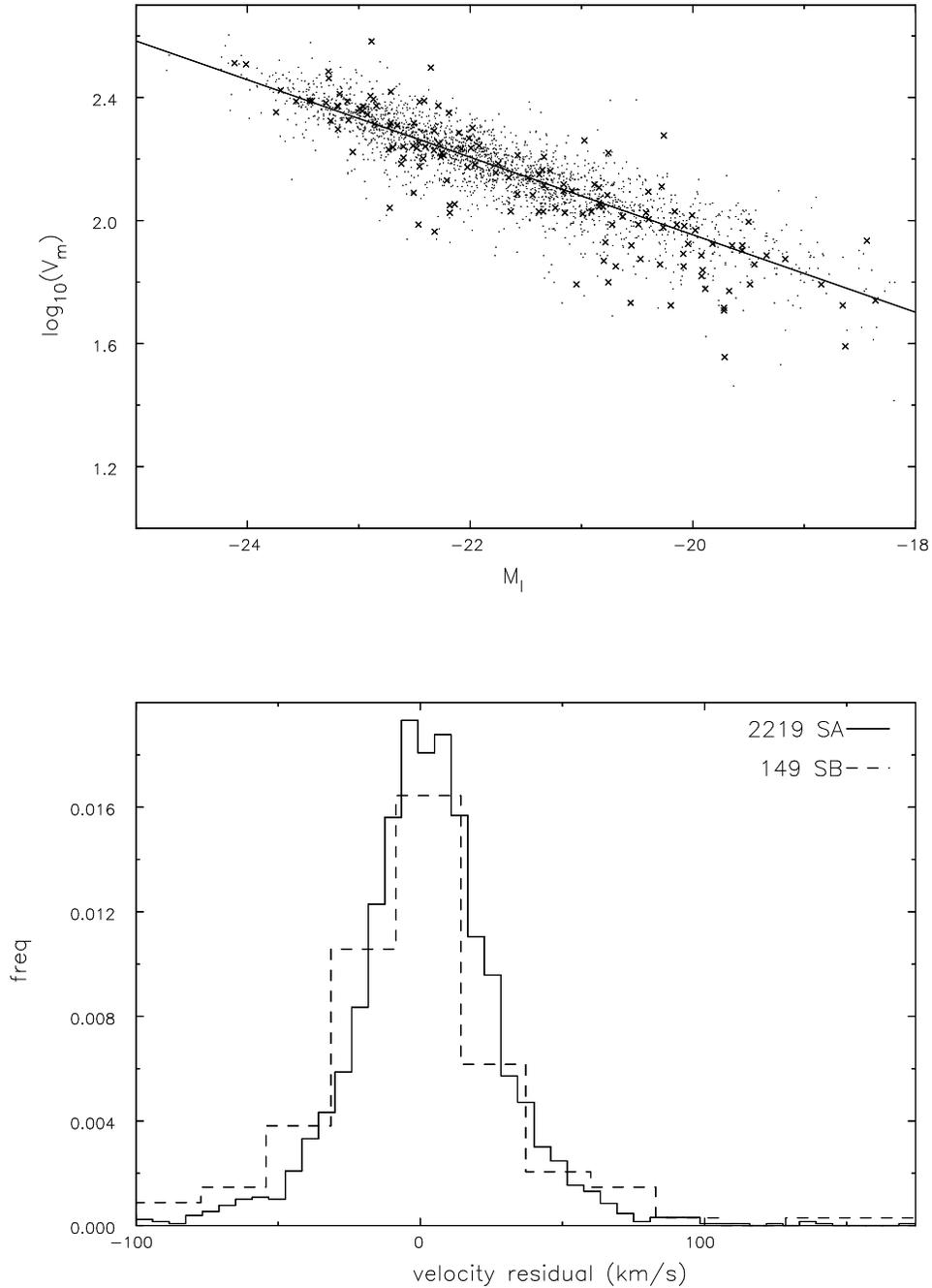,width=0.8\hsize,clip=,angle=0}}
\caption{The Tully-Fisher relation (a) and histograms of velocity residuals (b) for the sample collected by Mathewson \& Ford (1996).  Unbarred galaxies are plotted as points in (a), barred galaxies are marked by crosses and the slope and intercept of the fitted line in (a) are $-0.126$ \& $-0.566$ respectively.  The two histograms in (b) have been scaled so as to have equal (unit) area.}
\label{fig:TFR}
\end{figure}
\fi

It is often argued (\eg\ Courteau \& Rix 1999) that barred galaxies have lower density halos than do unbarred.  This prejudice stems from the paper by Ostriker \& Peebles (1973), who suggested that the only way to inhibit bar formation in a galaxy was to immerse the cool disk in a massive dynamically hot component.  Not only is this argument fallacious (Toomre 1981; Sellwood \& Evans 2000), but we here present evidence that bright barred galaxies have similar DM fractions as do their unbarred counterparts.

In Paper I, we argued against the hypothesis that barred galaxies have less DM than those of the unbarred family:  If the DM content varies continuously between maximum disk, fast bar, SB galaxies and massive halo SA galaxies, there should be many galaxies of intermediate dark matter content.  Any strong bars that may form in such galaxies would therefore be fiercely braked, as in our experiments.  Since no slow bars are known in HSB galaxies, we conclude that, either the distribution of dark matter is bimodal, or that all galaxies with moderate halo density have somehow avoided forming bars, both of which seem very unlikely, or that no galaxies are dark matter dominated.

Tidal triggering can induce a bar in a galaxy that is stable when isolated (\eg\ Noguchi 1987).  Such bars would be strongly braked if the target galaxy had been stabilized by a massive halo.  The absence of known slow bars again argues against massive halos, but only weakly; if the rate of bar-inducing tidal interactions is low, then the measured sample may be simply too small to include a slow case.

Empirical evidence against a systematic difference between barred and unbarred galaxies was presented by Bosma (1996) and more can be found in the data from Mathewson \& Ford (1996).  We use the apparent magnitudes in the I-band, recession velocities and $V_m$ given in their table, convert to absolute magnitudes assuming Hubble distances (for $H_0=60\;$km~s$^{-1}$~Mpc$^{-1}$) and plot the Tully-Fisher relation for the 2368 galaxies in their sample having recession speeds $> 1,000\;$km~s$^{-1}$ (to avoid absurdly inaccurate Hubble distances) in Figure~\ref{fig:TFR}(a).  The line is fitted to all the data, but the 2219 points are for ``unbarred'' galaxies and the crosses mark the 149 galaxies which Mathewson \& Ford designate as barred.\footnote{Their sample excluded galaxies designated as barred but, as always happens, bars were identified after the data were taken.  It is unclear whether these are typical bars, but since their barred fraction is extremely low, it seems likely that they flagged only the blatant, \ie\ strong, bars.}  The histograms in Figure~\ref{fig:TFR}(b) show the distributions of velocity residuals about the fitted line for the barred and unbarred galaxies separately (scaled so that the area under each histogram is unity), suggesting a small offset in the sense that the barred galaxies have slightly lower $V_m$ at a given brightness.  A Kolmogorov-Smirnov test indicates that there is 3.5\% probability that these two samples were drawn from the same parent distribution, suggesting a possibly significant difference.  However, the offset disappears if we discard all galaxies fainter than M$_{\rm I}\simeq-21$, indicating that it arises from the faint galaxies only, as is apparent in Figure~\ref{fig:TFR}(a).  We conclude that there is no evidence here for a deficiency in DM content, relative to the unbarred galaxies, in the (few) bright barred galaxies in the Mathewson \& Ford sample.  Further Tully-Fisher work with properly constructed samples of barred/unbarred galaxies to confirm this conclusion would be highly desirable.

We conclude that all bright HSB disk galaxies, barred or unbarred, are maximum disks.  Supporting evidence for this conclusion is reviewed by Sellwood (1999).

We also predict that any barred galaxy having a moderately dense halo should have a slow bar.  Prime candidates to test this prediction may be found amongst galaxies believed to have significant DM fractions in their inner regions: the low luminosity galaxies (\eg\ Persic \& Salucci 1988; see Sellwood 1999) and low surface brightness galaxies (LSBs, Bothun \etal\ 1997; de Blok \& McGaugh 1997).  Bars in these systems are less common, but not unknown.\footnote{The slight offset between the fainter barred and unbarred galaxies in Figure~\ref{fig:TFR}(a) is in the sense that the barred cases have a lower $V_m$ in relation to their luminosity, and presumably therefore a smaller DM fraction, than do the unbarred.  It is reasonable that bars {\it should\/} be found amongst those galaxies with more dominant disks.}  If a strongly barred low-luminosity or LSB galaxy has even a moderately dense DM halo, it should have a high value of $\s$.  Unfortunately, there are no reliable measurements of pattern speeds in such galaxies to test the prediction at this time.

\bigskip
We would like to thank Scott Tremaine for a number of insightful conversations and a critical read of the manuscript.  The comments of the referee, Lia Athanassoula, were also helpful.  This work was supported by NSF grant AST 96/17088 and NASA LTSA grant NAG 5-6037.  VPD acknowledges support of grant \# 20-50676.97 from the Swiss National Science Foundation for part of this work.

{

}

\appendix
\label{appa}

\section{Determination of an Equilibrium Distribution Function}

We here describe the creation of our equilibrium halo models in the presence of a massive disk.  The procedure is identical to that followed by Raha \etal\ (1991), but has not previously been explained in detail in a published article.  Because the equilibrium generated by this procedure is exact, there is no need to compute the evolution of the halo while it adjusts to an equilibrium from an approximate set-up (\eg\ Barnes 1992; Hernquist 1993).

We adopt the iterative approach to finding a DF first proposed by Prendergast \& Tomer (1970) and developed by Jarvis \& Freeman (1985) for two component systems and Kuijken \& Dubinski (1995) for a three component system.  Unlike these authors, however, we solve for the gravitational potential using the same numerical procedure that is used in the simulations, thereby incorporating any numerical idiosyncrasies of the potential determination into the solution for the DF; this strategy ensures that the particle distribution is in perfect equilibrium at the outset.

We first choose a functional form for the DF of the halo
\begin{equation}
f = C {\cal F}(I_1, I_2, ...),
\end{equation}
where $C$ is a normalization constant and $\cal F$ can be more or less any reasonable function of the classical isolating integrals, $I_n$.  In our axisymmetric potential, these are $E$ and $J_z$.  The form of $\cal F$ adopted determines the density profile and shape of the resulting halo; functions of $E$ alone tend to produce almost spherical halos (the disk makes them slightly oblate), while adding a dependence on $J_z$ generally produces more strongly spheroidal halos.

A first approximation to the halo density $\rho_1(R,z)$ can be determined from \begin{equation}
\rho_1(R,z) = \int f \, d^3\bv 
\end{equation}
using any reasonable initial guess for the gravitational potential $\Phi_1(R,z)$.  We assign mass to the grid to represent the smooth function $\rho_1(R,z)$, add the mass distribution of a smooth disk and solve for a new gravitational field $\Phi_2(R,z)$.  We then determine $\rho_2(R,z)$ using the improved potential in (A2), and iterate until the potential distribution converges.  The value of $C$ can be adjusted at each iteration step to drive the solution to the desired halo mass.  We find that the solution converges rapidly and that 10 iterations are usually ample.  

Note that although the halo density profile, and therefore the net rotation curve, cannot be specified in advance, the rapid convergence permits many models to be explored (for different mass ratio, truncation radius, choice of $\cal F$, \etc)\ from which one having the desired properties may be selected.

\section{Quiet Start Procedures}
\label{quietstart}

Since an $N$-body simulation amounts to a numerical solution of the coupled collisionless Boltzmann and Poisson equations by the method of characteristics, it is clearly desirable to select the characteristics to be followed with care.  Selecting particles at random from a DF (\eg\ Hernquist, Spergel \& Heyl 1993; Kuijken \& Dubinski 1995) leads to $\sqrt N$-type variations in the number of particles generated in any given range of the integrals; in effect the model will have the dynamical properties of one with a DF slightly different from that intended, which has many disadvantages, especially when attempting comparisons with theoretical work.  The following quiet start procedures lead to much higher quality simulations (and are also more efficient); every part has been described in some other publication, but for ease of reference we summarize them here.

Strategies for the optimal selection of points are exactly analogous to those for the selection of abscissae in the numerical evaluation of multi-dimensional integrals.  In that case, accuracy is improved whenever the dimensionality can be reduced by analytic integration over some of the coordinates.  In our problem, we know the DF to be independent of orbital phases, since they must be uniformly populated in any equilibrium model.  Note that, except when the DF is expressed in terms of actions, the density of particles in the sub-space of the integrals is not simply given by the DF; it needs to multiplied by a ``density of states'' function, which is the phase-space volume per unit interval of $E$ and $J_z$ (Binney \& Tremaine 1987, \S4.4.5)

For a razor-thin disk, the density of particles having energy $E$ and angular momentum $J_z$ is (Sellwood \& Athanassoula 1986)
\begin{equation}
{\cal N}_{\rm disk}(E,J_z) = 2\pi f(E,J_z) \tau(E,J_z), 
\end{equation}
where $f$ is the usual phase space density and $\tau(E,J_z)$ is the period of one complete radial oscillation of a particle with the given $(E,J_z)$.  The latter generally has to be determined numerically.  For a sphere with 
$f(E,L)$ 
\begin{equation}
{\cal N}_{\rm sphere}(E,L) = 8\pi^2 L f(E,L) \tau(E,L), 
\end{equation}
(Binney \& Tremaine 1987, problem 4-22), while, for a spheroid with 
$f(E,J_z)$, 
\begin{equation}
{\cal N}_{\rm spheroid}(E,J_z) = 4\pi^2 f(E,J_z) S(E,J_z)  
\end{equation}
(Sellwood \& Valluri 1997).  In this last formula, $S(E,J_z)$ is the cross-sectional area in $(R,z)$ of the bounding torus in the meridional plane (Binney \& Tremaine 1987, \S3.2.1) and is easily evaluated numerically for arbitrary potentials.

We proceed by slicing accessible $(E,J_z)$ space into a number, $j=n_En_J$, of small areas in such a way that the integral of the DF over each area encloses a fraction $1/j$ of the total active mass; we generally choose $n_E \gg n_J$.  We then select a point within each of these areas to determine the $(E,J_z)$ values for an orbit.  We avoid a regular raster of such points in $(E,J_z)$ space while maintaining the desired smoothness, as follows:  For every slice in energy, we choose $n_J$ equal spaced values of $J_z$ from the distribution of ${\cal N}|_E$, with the first value only determined as a random sub-fraction, and then select an $E$ value within each area at random from the distribution of $\int {\cal N}(E,J_z)dJ_z$.

Each $(E,J_z)$ pair selected in this way specifies an orbit and we must next choose phases to determine both the initial position and velocity components of the particles.  In contrast to the selection of integrals, experience suggests that the behavior of the model is much less sensitive to the manner in which some orbital phases are selected.  In general, random selection from the appropriate distribution is adequate, although it is easy to improve upon random when desired for a particular application.  Two examples are: Sellwood (1983) found it desirable to space several otherwise identical particles equally around a ring when searching for small-amplitude non-axisymmetric instabilities and Sellwood (1997) was able to quieten radial pulsations of a stable spherical model by spacing particles equally in radial phase.

In a razor-thin disk, or in a sphere, the orbit lies in a plane in which particles oscillate between peri- and apocenter with full period $\tau(E,L)$.  The radial phases must be uniformly distributed, but the probability of selecting a particular radius varies inversely with the (non-uniform) radial speed.  It is easiest to select a fraction of the radial period and integrate the orbit (usually numerically) for this time to determine the radius.  The radial and azimuthal speeds are completely determined (except for the sign ambiguity of the radial speed) by the selected values of $E$, $J_z$ and $r$.  The azimuthal phase and, for a sphere, the orientation of the plane can be selected in a straightforward manner.

In spheroidal models, as here, the two classical integrals confine the particle to a torus in real space.  When the desired DF does not depend upon a third integral, the probability density distribution for any given orbit is uniform in the meridional plane within the boundaries of the torus and selection of $(R,z)$ pairs is straightforward.  The values of $E$ and $J_z$ almost determine the velocity components at the chosen position -- all that remains is to direct the velocity component in the meridional plane, which should be uniformly distributed in angle.

\section{Halo Angular Momentum}
\label{apphaloangmom}

In a general axisymmetric system, the dependence of the density $\rho$ on $J_z$ is only through the {\it even\/} part of the DF (Lynden-Bell 1962).  Thus angular momenta about the symmetry axis $z$ can be reversed at random without affecting the equilibrium of the system.  Changing the net angular momentum may, however, alter the stability of the system; in particular, Kalnajs (1977) has cautioned that a discontinuity in the DF at $J_z = 0$ can aggravate instabilities.  We therefore adopt a scheme which ensures a smooth DF.

If a prograde halo is desired, we define $p_\alpha(x)$, where $x=-J_z/J_{z,max}$, to be the probablity of changing the sign of $J_z$ of a retrograde particle in a halo in which the maximum possible angular momentum at the truncation energy is $J_{z,max}$.  (To generate a retrograde halo, we flip only particles having positive $J_z$ with probablity $p_\alpha(x)$, where now $x=J_z/J_{z,max}$.)

In order to make the DF continuous at $J_z=0$, we require $p_\alpha \rightarrow 0$ as $x \rightarrow 0$.  By extending the definition of $p_\alpha(x)$ to be an odd function when $x<0$, we can write the new distribution function as $\tilde{f}(E,J_z) = f(E)[1-p_\alpha(x)]$.

We adopted the shifted, normalized Fermi function and its inverse, which both have the desired properties:
\begin{equation}
p_\alpha(x) = \left\{ \begin{array}{ll}
-\left[(e^{\alpha x} + 1)^{-1} - \frac{1}{2}\right] / \left[(e^{-\alpha} + 1)^{-1} - \frac{1}{2}\right] & \ {\rm large \ } L_{z,h} \\
-\frac{1}{\alpha} \ln \left( \left\{x\left[(e^{-\alpha} + 1)^{-1} - \frac{1}{2}\right] + \frac{1}{2}\right\}^{-1} - 1 \right) 
& \ {\rm small \ } L_{z,h}
\end{array} \right.
\label{eq:flipe}
\end{equation}
When $\alpha = 0$, both functions are $p_0(x)=x$, which leads to a certain total $L_{z,h}$.  If the desired total $L_{z,h}$ is greater (smaller) than this value, we use the large (small) expression and adjust $\alpha$ to yield the desired net angular momentum.

In order to generate a radial variation in the net rotation, we make the probability a function of energy as follows:
\begin{equation}
p_d(x) = -N(E) \frac{\left( e^{\alpha x} + 1 \right)^{-1} - \frac{1}{2}}
{\left( e^{-\alpha} + 1 \right)^{-1} - \frac{1}{2}}
\label{eq:flipf}
\end{equation}
with
\begin{equation}
N(E) = \left\{ \begin{array}{ll}
1 & E \leq E_1 \ \equiv \ E_{\rm min} + (E_{\rm max} - E_{\rm min})\frac{1}{d} 
\\
2 s^3 - 3 s^2 + 1 & E_0 > E > E_1 \\
0 & E \geq E_0 \ \equiv \ E_{\rm min} + (E_{\rm max} - E_{\rm min})(\frac{1}{d} 
+ \frac{1}{10}) \\
\end{array} \right.
\end{equation}
Here $s = (E - E_1)/(E_0 - E_1)$, while $E_{\rm min}$ and $E_{\rm max}$ are constants of the system.  We fix $\alpha = 30$ to ensure the inner halo rotates strongly and vary $d$ to adjust the energy at which strong rotation gives over to no rotation.

\end{document}